\documentclass{aa}  

\usepackage{graphicx}
\usepackage{txfonts}
\usepackage{booktabs}
\usepackage{siunitx}
\usepackage{tabularx} 
\usepackage{amsmath}
\usepackage{float}
\usepackage{dblfloatfix}
\usepackage{placeins}
\begin{document}

   \title{Linear Spectropolarimetry of the 2011 Eruption of the Recurrent Nova T Pyx}
   \subtitle{Evidence for a Multi-component Ejecta Geometry}

   \author{A. Lamy-Rundstadler
          \inst{1}
          \and
          A. Ederoclite\inst{2}
          }

   \institute{Universidad Internacional de Valencia (VIU), C/Pintor Sorolla 21, 46002, Valencia, Spain\\
              \email{agmaj.lamy@gmail.com}
         \and
             Centro de Estudios de Física del Cosmos de Aragón (CEFCA), Unidad Asociada al CSIC, Plaza San Juan 1, 44001 Teruel, Spain\\
             \email{aederocl@cefca.es}
             }

   \date{received ; accepted}

 
  \abstract
   {The geometry of classical nova ejecta is commonly investigated through direct imaging and optical or radio interferometry once the expanding material becomes spatially resolved. These techniques reveal that the ejecta geometry is often highly asymmetric. Linear spectropolarimetry provides an alternative means of probing ejecta asymmetries during the early phases of an eruption, when the source remains spatially unresolved. However, only a handful of novae have been monitored with sufficient temporal coverage to follow the evolution of their intrinsic polarization.}
   {We investigate the temporal evolution of the intrinsic polarization of the 2011 eruption of the recurrent nova T Pyxidis and use its continuum and line-polarization signatures to constrain the geometry and kinematic structure of the ejecta.}
   {We analyzed nine epochs of VLT/FORS2 optical linear spectropolarimetry obtained between 24 and 70 days after discovery, from shortly before optical maximum to the early decline. The instrumental and interstellar polarization contributions were removed, with the latter estimated from the depolarization observed at the rest wavelengths of the Balmer emission components. We examined the continuum and the velocity-dependent intrinsic polarization across the H$_\beta$, H$_\gamma$, and H$_\delta$ P Cygni profiles.}
   {The intrinsic continuum polarization generally remains below 1\% and is predominantly aligned at a position angle of approximately 95°, except near optical maximum, when an orthogonal component at approximately 185° temporarily dominates. The blueshifted Balmer absorptions display stronger polarization, frequently reaching 2\%, associated with excursions and loops in the Q-U plane. At least two distinct velocity-dependent polarization components are detected: the higher-velocity absorption is preferentially associated with the 95° orientation, whereas the lower-velocity absorption generally traces the 185° orientation. The polarization peaks shift quasi-linearly towards higher blueshifted velocities with time in all three Balmer lines.}
   {The observations reveal a structured, multi-component ejecta geometry whose polarization properties evolve with velocity and optical depth. The dominant continuum polarization and the lower-velocity absorption imply a projected structural axis close to 5°, consistent with the equatorial or toroidal morphology inferred from spatially resolved observations, while the orthogonal component may trace bipolar ejecta. These results indicate that the large-scale equatorial and bipolar structures observed at later stages were already established during the first months of the eruption.}


   \maketitle
%

\section{Introduction}


Classical novae are thermonuclear explosions occurring on the surface of a white dwarf accreting mass from a less-evolved Roche-filling companion. The explosion is triggered when the accreted material reaches the ignition pressure to start thermonuclear reactions (for a review, see \citealt{Bode+Evans}).

The thermonuclear explosion occurs on the surface of the white dwarf, and the evolution of the ejected material is driven primarily by the cooling of the thermonuclear ashes, the changing optical and geometrical depth of the ejecta, and their interaction with the immediate surroundings. Throughout this evolution, the spectrum is dominated by hydrogen lines, whose profiles evolve characteristically from P Cygni\footnote{\citet{Mason+2018} argue that the term P Cygni is inappropriate here, as the velocity profile is incompatible with a wind. We preserve the "P Cygni" name for compatibility with the rest of the literature.} to flat-topped, reflecting a transition from higher to lower densities \citep{Williams1992,Williams2012}. Beyond the Balmer series, the most prominent lines in optical spectra are typically either \ion{Fe}{II} 
or He/N. This distinction in the dominant line species motivated the classification scheme developed by \citet{Williams1992}, though \citet{Aydi+2020} later argued that the two classes are an artifact of observational bias rather than a physically distinct dichotomy.

When the ejecta become spatially resolved, various features such as polar caps or equatorial rings can be observed (e.g. \citealt{Santamaria+2025}). Deviations from spherical symmetry have been directly measured with interferometry, both in the optical \citep{Chesneau+2011} and, more extensively, in radio (e.g. \citealt{Chomiuk+2014Natur}). These asymmetries have also been explored theoretically, with works such as \citet{Figueira+2018} modelling the interaction of the ejecta with an accretion disc and a companion star.

In this phenomenon, the white dwarf is not destroyed and accretion is resumed (see \citealt{Shara+2018}). This makes it possible for a nova to explode again, once the ignition pressure is reached again. The time between two consecutive nova explosions in the same binary system is called ``recurrence
time''. Typical recurrence times are estimated to be in the order of $10^4$\,years. 
In a few cases, the recurrence time is much shorter and more than one explosion is observed in a system. We refer to these explosions as ``recurrent novae'' (see \citealt{Schaefer2010}). Most known recurrent novae occur in symbiotic systems, where the secondary is a 
red giant and accretion proceeds via wind capture. However, two recurrent novae are thought to occur in systems resembling classical novae: IM\,Nor \citep{Patterson+2022} and T\,Pyx.

The first recorded explosion of T\,Pyx occurred in 1890. Since then, five other eruptions have been observed (1902, 1920, 1944, 1966, and 2011). The last eruption of T\,Pyx was discovered on 14 April 2011 \citep{discovery_IAUC}. Given the large gap between the 1966 and the 2011 explosion (see \citealt{Gilmozzi+Selvelli2007}, \citealt{Selvelli+2008}, and \citealt{Schaefer+2010}), this last explosion was followed in several wavelengths (e.g. \citealt{Surina+2014}).

Linear spectropolarimetry provides a powerful means of probing the three-dimensional geometry of nova ejecta at epochs when the source remains spatially unresolved. In a perfectly spherical configuration, the polarization vectors produced by scattering processes cancel out when integrated over the projected surface of the source, resulting in zero net polarization. Therefore, any measurable linear polarization directly signals a departure from spherical symmetry. In novae, continuum polarization is primarily produced by Thomson scattering of photospheric photons by free electrons in the ionized ejecta. If the electron-scattering region is aspherical (such as an oblate or prolate outflow, an equatorial density, or a bipolar structure), the incomplete cancellation of the polarization vectors leads to a detectable net signal.

Spectral lines provide additional geometric diagnostics. Line photons generally form over a more extended volume than the underlying pseudo-photosphere and often experience fewer scatterings than continuum photons. As a result, emission lines tend to dilute the polarized continuum, producing depolarization signatures across the line profile. In contrast, line polarization can arise when absorbing or scattering material is distributed asymmetrically in front of an already polarized photosphere formed by Thomson scattering. In this case, the line effect is a direct consequence of the non-spherical distribution of ejecta along the line of sight, selectively blocking or redistributing polarized flux, and producing characteristic signatures in the Stokes parameters. The analysis of polarization as a function of wavelength and its representation in the Q-U plane allows one to distinguish between simple axisymmetric geometries and more complex or evolving morphologies. However, the observed polarization signal may also include contributions from interstellar polarization (ISP) along the line of sight. Correctly estimating and removing this ISP component is therefore essential in order to recover the intrinsic polarization of the ejecta and reliably interpret their geometry.

Because nova ejecta are known to develop equatorial rings, polar caps, and clumpy substructures at later resolved stages, linear spectropolarimetry offers a unique opportunity to diagnose these asymmetries during the earliest phases of expansion, when the geometry is otherwise inaccessible.

\section{Observations and Data Reduction}

%
%
%
%

\subsection{Observations}

Spectropolarimetric data were obtained using the FOcal Reducer and low-dispersion Spectrograph (FORS2) mounted at the Cassegrain focus of Unit Telescope 1 (Antu) of the Very Large Telescope (VLT) at the European Southern Observatory (ESO), Paranal, Chile. In PMOS mode, FORS2 operates as a dual-beam optical spectropolarimeter that measures linear polarization with a halfwave retarder plate that rotates the plane of polarization before a Wollaston prism separates the light into two orthogonally polarized beams (the ordinary (o) and extraordinary (e) beams). These two beams are recorded simultaneously on the detector using a striped PMOS slit mask with a slit width of 0.5'' to avoid overlapping the spectra. A series of spectra is obtained at different rotation angles of the retarder plate, $\theta_i = i \times 22.5^{\circ}$ with $0 \le i \le 3$, which is generally sufficient to reconstruct the linear polarization state of the source.

\begin{figure}
\centering
\includegraphics[width=1.0\linewidth]{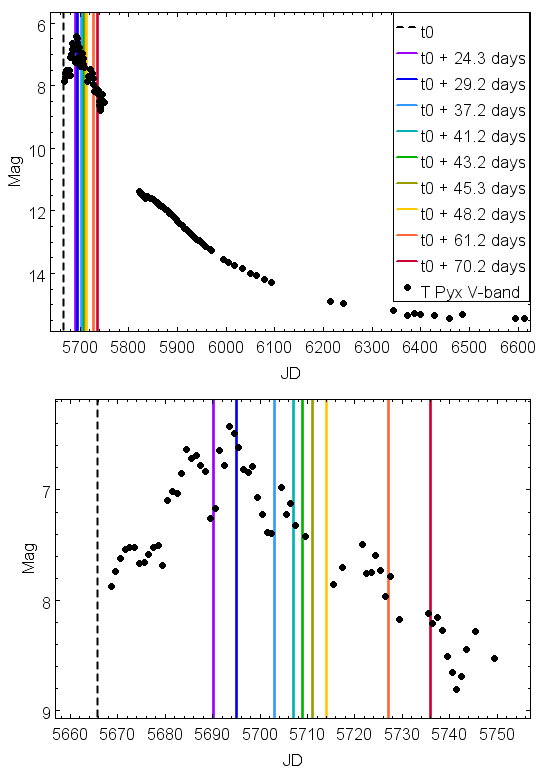}
\caption{Light curve of T Pyxidis (from “The STONY BROOK/SMARTS Atlas of (mostly) Southern Novae” 
\citealt{Walter+2012}) and position of the 
observations.}
\label{fig:lightcurve}%
\end{figure}

The eruption of T Pyxidis was detected with a visual magnitude of 13.0 on April 14, 2011, at 14:29 UT (JD = 2455665.79), hereafter referred to as $t_0$ (Waagan et al. 2011). Linear spectropolarimetric observations were conducted over nine different epochs from May 9th ($t_0+24.3\,\mathrm{d}$), to June 23rd 2011 ($t_0+70.2\,\mathrm{d}$), spanning the light-curve maximum and the onset of the decline. Fig.~\ref{fig:lightcurve} highlights the epochs of observation over the optical light curve of T Pyx.

The optical polarimeter in FORS2 was configured with the standard resolution collimator, providing a plate scale of 0.2 arcsec/pixel. The object was placed in the central slit of the PMOS focal mask, close to the optical axis of the instrument. The 1200B grism was used, providing a wavelength coverage of 3670--5128 \AA\ and a spectral dispersion of 0.36 \AA\,pixel$^{-1}$. 
Two cycles of observation were taken at each epoch to check the consistency of the results. Several polarized standard stars and unpolarized standard stars were also observed at each epoch to calibrate instrumental polarization and angle offset. 
A preliminary inspection of the data was performed to identify and discard some frames with poor image quality. The list of remaining 
observations including the dates and exposure times for each epoch is presented in Table~\ref{tab:time-expt.time}. 

\begin{table}
  \caption{2011 Spectropolarimetric Observation Log}
  \label{tab:obslog}
  \centering
  \small
  \begin{tabularx}{\columnwidth}{c l S[table-format=4.2] S[table-format=2.2] >{\centering\arraybackslash}X}
    \toprule
    {Obs \#} & {Date (2011)} & {JD-2450000} & {$t-t_0$ (days)} & {Exp. time per angle} \\
    \midrule
    1 & May 9th  & 5690.06 & 24.27 & $2\times5\,\mathrm{s}$ \\
    2 & May 14th & 5694.99 & 29.20 & $2\times2\,\mathrm{s}$ \\
    3 & May 21st & 5702.96 & 37.17 & $2\times20\,\mathrm{s}$ \\
    4 & May 25th & 5706.97 & 41.18 & $2\times6\,\mathrm{s}$ \\
    5 & May 27th & 5708.98 & 43.19 & $2\times15\,\mathrm{s}$ \\
    6 & May 30th & 5711.09 & 45.30 & $2\times25\,\mathrm{s}$ \\
    7 & June 1st & 5713.97 & 48.18 & $2\times45\,\mathrm{s}$ \\
    8 & June 14th& 5726.98 & 61.19 & $2\times40\,\mathrm{s}$ \\
    9 & June 23rd& 5736.00 & 70.21 & $2\times20\,\mathrm{s}$ \\
    \bottomrule
  \end{tabularx}
  \label{tab:time-expt.time}
\end{table}

\subsection{Data Reduction}

Data reduction was carried out using ESOReflex and the dedicated FORS\_PMOS pipeline v2.11.5, which includes bias subtraction, flat fielding, correction of spatial distortions, sky subtraction, extraction and wavelength calibration of the o and e spectra. 
The differences of ordinary and extraordinary fluxes are first normalized:

\begin{equation}
F(\theta_i) = 
\frac{f_o(\theta_i) - f_e(\theta_i)}
     {f_o(\theta_i) + f_e(\theta_i)}.
\end{equation}

The Stokes parameters $Q$ and $U$ are obtained using a discrete Fourier transform applied to the measurements taken at several half-wave plate angles:

\begin{align}
Q &= \frac{2}{N} \sum_{i=0}^{N-1} F(\theta_i)\cos(4\theta_i)\\
U &= \frac{2}{N} \sum_{i=0}^{N-1} F(\theta_i)\sin(4\theta_i)
\end{align}

where $\theta_i = i \times 22.5^{\circ}$ is the angle of the half-wave plate and $N$ is the total number of positions used.

The pipeline also provides the degree of linear polarization $P$ and the position angle $\theta$, based on the observed Stokes parameters, and calculated using equations \ref{eq:Degree} and \ref{eq:Angle}.

\begin{align}
P &= \sqrt{Q^{2} + U^{2}} 
\label{eq:Degree}\\
\theta &= \frac{1}{2}\operatorname{atan2}(U,Q)
\label{eq:Angle}
\end{align}

Pairs of individual data sets taken during the same night were combined (using the mean value $\bar{X}$ - weighted by the inverse square of the associated errors of the Stokes parameters) to improve the signal-to-noise ratio (SNR).

\begin{equation}
\bar{X} = 
\frac{\displaystyle \sum_i \frac{X_i}{\sigma_{X_i}^2}}
     {\displaystyle \sum_i \frac{1}{\sigma_{X_i}^2}}.
\end{equation}

where the index \(i \in \{1,2\}\), \(X\) denotes either of the Stokes parameters \(Q\) or \(U\), and \(\sigma_X\) is the corresponding uncertainty.
To further reduce noise without losing the signature of the line effect, the spectra were smoothed using a weighted moving average with a 1.8\,\AA\ window.

\subsection{Data Corrections}

The spectropolarimetric data were corrected for instrumental and interstellar polarization. Although interstellar polarization mainly produces a translation in the $Q/U$ plane, it modifies the derived polarization degree and position angle through the non-linear transformation from the Stokes parameters, making its removal necessary to recover the intrinsic polarization of the ejecta. The ISP was estimated assuming complete depolarization of the Balmer recombination lines. The validity of this assumption is supported by the minimal dispersion of the Stokes parameters measured at the Balmer rest wavelengths throughout the observing campaign. Full details of the correction procedure are presented in Appendix~\ref{app:corrections}.

\section{Results}

\begin{figure*}
    \centering
    \includegraphics[width=\textwidth]{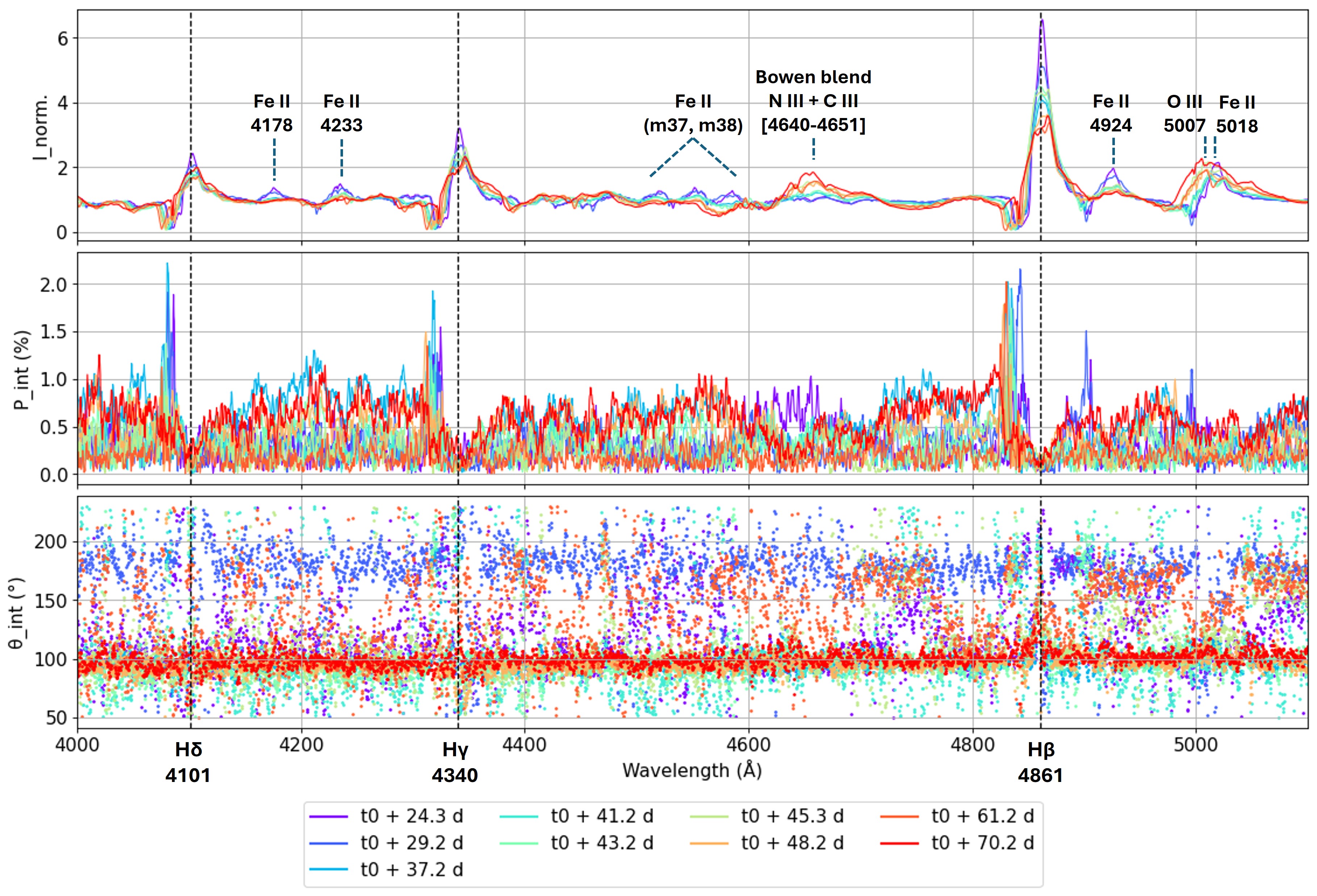}
    \caption{Multi-epoch spectropolarimetric comparison of the three
Balmer lines H$_\beta$, H$_\gamma$, and H$_\delta$.
Top panel: normalized intensity spectra.
Middle panel: intrinsic polarization degree $P_{\rm int}(\lambda)$.
Bottom panel: intrinsic polarization angle $\theta_{\rm int}(\lambda)$.
Each color corresponds to a different observing epoch. The vertical dashed lines indicate the rest wavelengths of the Balmer transitions.}
\label{fig:results}
\end{figure*}

The results are presented in Fig.~\ref{fig:results}.
The intensity spectrum covers the three Balmer lines H$_\beta$, H$_\gamma$, and H$_\delta$, which clearly display the evolution of P-Cygni profiles. The continuum polarization remains mostly below 1\%. At the rest wavelength of each Balmer line, the degree of polarization reaches a minimum, whereas spikes of up to 2\% are observed on the absorption side of the P-Cygni profiles. The polarization angle is distributed mainly around 95° and 185°.

\begin{figure*}
    \centering
    \includegraphics[width=\textwidth]{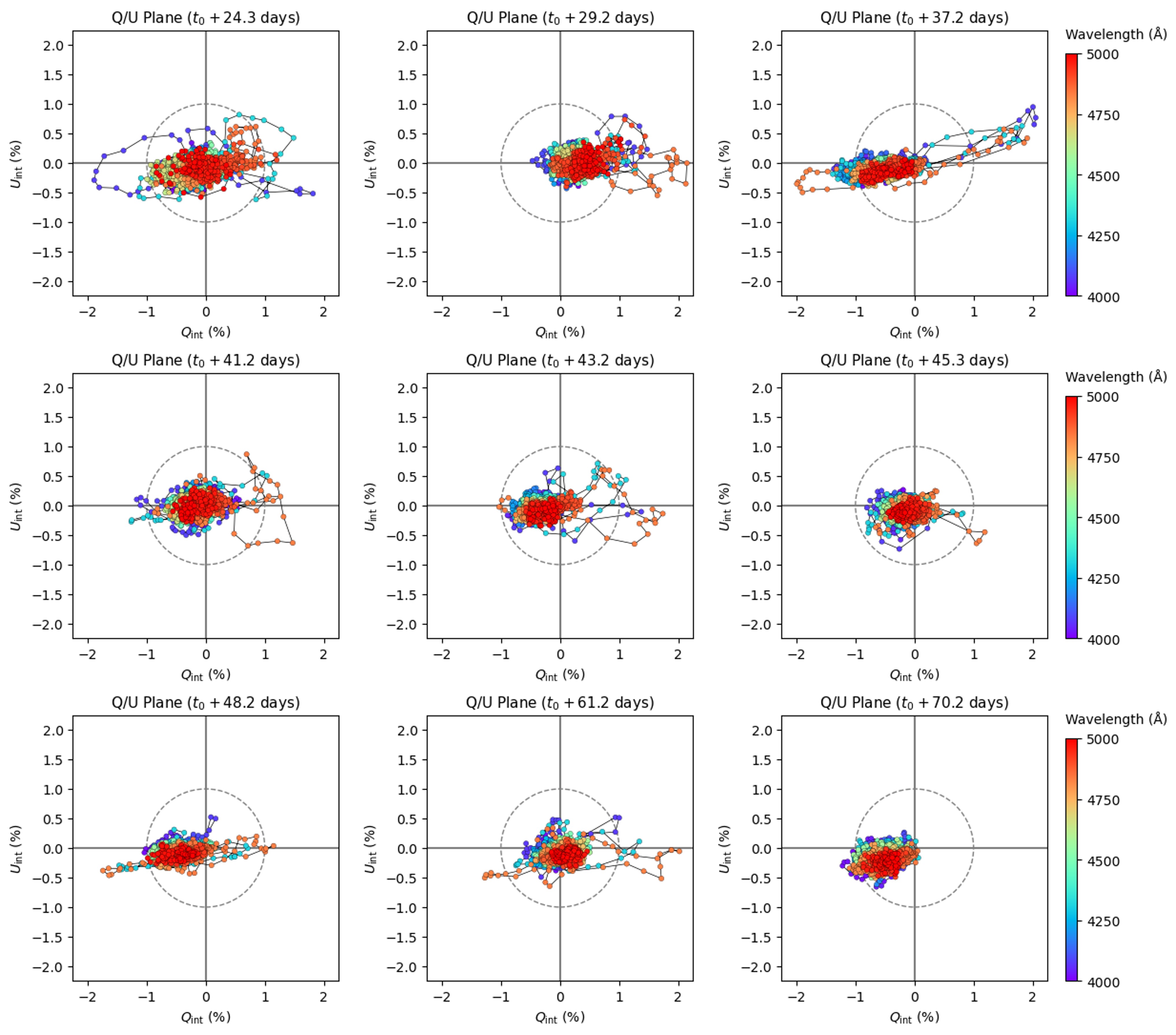}
    \caption{Intrinsic spectropolarimetric data of T~Pyx in the Q/U plane for each observing epoch. Each panel corresponds to one epoch of the 2011 outburst. The colour scale represents wavelength across the observed spectral range (4000-5000\,\AA). Points belonging to the same spectral feature are connected to guide the eye and illustrate the polarization excursions across the line profiles. The dashed circles indicate levels of constant polarization degree of 1\%.}
\label{fig:QU tous}
\end{figure*}

In the Q/U plane, after ISP correction, some polarization loops clearly cross into the opposite quadrant with respect to the continuum, indicating an intrinsic position angle change of 90° (a rotation of 180° in the Q-U space corresponds to a real variation of 90° in the sky). The data for each epoch are displayed in the Q-U plane in Fig.~\ref{fig:QU tous}.

%

\subsection{Spectral Evolution}

Normalized spectra (Fig.~\ref{fig:results}) illustrate the spectroscopic evolution of T~Pyx during the 2011 eruption. At the earliest epochs, several \ion{Fe}{II} 
emission lines are clearly visible, together with
strong Balmer lines (H$_\delta$ $\lambda4101$, H$_\gamma$ $\lambda4340$, H$_\beta$ $\lambda4861$). These \ion{Fe}{II}
features are characteristic of the optically thick stage of the ejecta and progressively weaken as the ionization increases and the envelope becomes more transparent.
In T~Pyx, the  \ion{Fe}{II} 
signatures fade and become marginal after roughly $\sim40$-45 days from discovery, marking the transition toward a higher-ionization regime. This behavior is consistent with the spectroscopic evolution described by \cite{Shore+2011} and interpreted
in the framework of the evolutionary sequence proposed by \cite{Aydi+2024}, in which novae evolve from an early He/N phase to an   
\ion{Fe}{II}-dominated stage near optical maximum, followed by a later He/N phase as the ejecta expand and their optical depth decreases.

We also observe the development of a broad emission structure near 4640-4651\,\AA, commonly identified as the Bowen blend and composed primarily of \ion{N}{III} 
and 
\ion{C}{III}
fluorescence lines. The emergence of this feature signals the onset of the post-maximum He/N phase, when the ejecta become progressively optically thin and photoionization from the hot central source increasingly dominates the excitation conditions.

\subsection{Global Polarimetric Evolution}

The intrinsic polarization exhibits significant variability throughout the observing campaign (Fig.~\ref{fig:results}). The continuum polarization generally remains below 1\%, although both its degree and position angle vary from one epoch to another. In contrast, the spectral lines display much stronger polarimetric signatures. The Balmer absorption components frequently reach polarization levels above 1\%, locally exceeding 1.5\%, whereas the corresponding emission components are systematically associated with a decrease in polarization, consistent with partial depolarization.

Despite this variability, the polarization position angle is not randomly distributed. Instead, it is preferentially clustered around two orientations, approximately 95$^\circ$ and 185$^\circ$, hereafter referred to as PA1 and PA2, respectively (with the angles shown in an unwrapped representation; PA2 = 185$^\circ$ is equivalent to 5$^\circ$ modulo 180$^\circ$). The relative contribution of these two preferred orientations changes from one epoch to another. Around optical maximum ($t_0+29.2$ d), the continuum is weakly polarized ($\lesssim0.5\%$) and is predominantly aligned with PA2, whereas the latest observation ($t_0+70.2$ d) is characterized by a continuum polarization ($\gtrsim0.5\%$) aligned with PA1. Most of the remaining epochs display intermediate behaviours, with the continuum showing contributions from both preferred orientations.

Epoch~1 deserves particular attention. Besides the polarization associated with the \ion{Fe}{II} 
$\lambda4924$ absorption, it is the only observation displaying significant polarization across the Bowen blend region 
(\ion{N}{III}/\ion{C}{III}), with a polarization degree locally exceeding 0.5\%. This signature disappears in all subsequent epochs, following the spectroscopic evolution of the nova toward the post-maximum He/N phase.

\subsection{Balmer Line Profiles}

\begin{figure}
\centering
\includegraphics[width=1.0\linewidth]{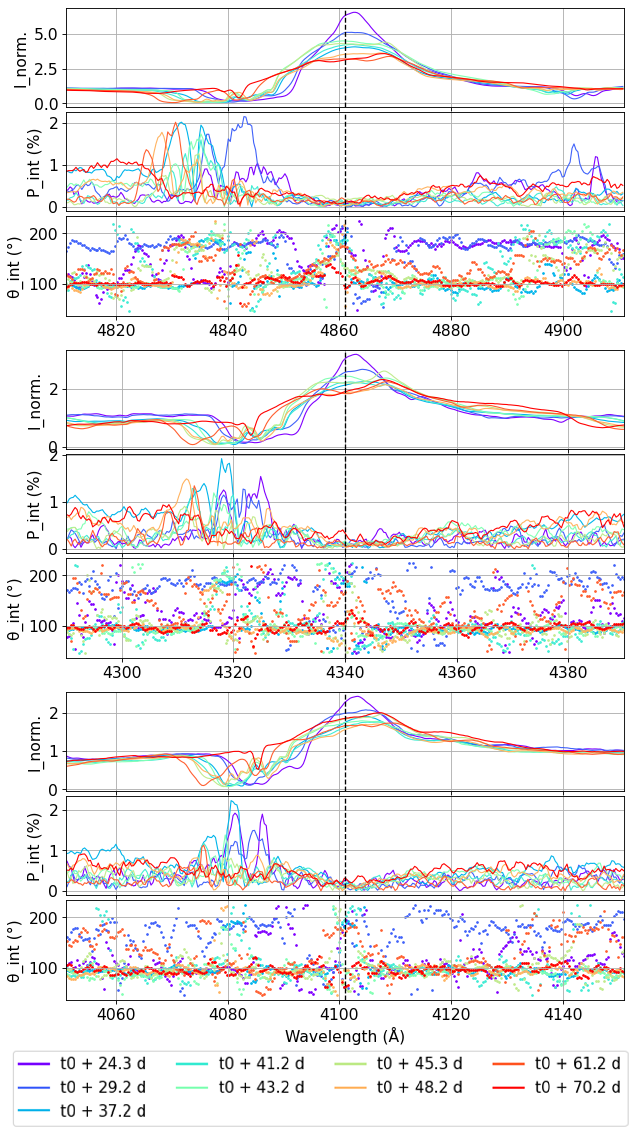}
\caption{Multi-epoch spectropolarimetric comparison of the Balmer lines H$\beta$, H$\gamma$, and H$\delta$. Each block corresponds to one Balmer transition (from top to bottom: H$\beta$, H$\gamma$, H$\delta$). For each line, the three panels show the normalized intensity profile, the intrinsic polarization degree $P_{\rm int}(\lambda)$, and the intrinsic polarization angle $\theta_{\rm int}(\lambda)$, respectively. Different colours represent the various observing epochs. The vertical dashed lines indicate the rest wavelength of each Balmer transition.}
\label{fig:multiepoch-Hbeta}%
\end{figure}

The spectral coverage of FORS2 enables a direct comparison of the spectropolarimetric properties of three Balmer lines (H$_\beta$, H$_\gamma$, and H$_\delta$) within the same spectrum and at identical resolution. Common features are observed in the P-Cygni profiles of these three lines, as shown in Fig.~\ref{fig:multiepoch-Hbeta}.

The early spectra show narrow emission features characterized by relatively low expansion velocities (of the order $\sim500$ km\,s$^{-1}$). The predominance of low projected velocities at these stages may indicate a significant contribution from material located near the projected limb of the expanding ejecta, where the radial velocity component along the line of sight is minimal. About one week after maximum light, a broader emission component emerges, roughly twice the width of the initial P-Cygni profile. The absorption component of the original P-Cygni profile remains superimposed on this broader emission. Such a spectral evolution has long been recognized \citep{McLaughlin1943,Gallagher+1978} 
and was recently confirmed to be nearly universal among novae \citep{Aydi+2020}.

On the absorption side, the P-Cygni profiles reveal multiple layers with distinct characteristics. Later, additional components appear at higher velocities. This phenomenon has been documented in various nova studies \citep{Hutchings1970,Shore+2011}, and its occurrence seems to coincide with secondary flares in the optical light curve \citep{Csak+2005,Tanaka+2011}. 
A marked increase in polarization is observed in the absorption troughs of the P-Cygni profiles, consistent with the differential polarization effect induced by selective absorption \citep{McLean1979}. 
In this scenario, polarization peaks naturally align with absorption maxima, as the selective blocking of forward-scattered (less polarized) light enhances the relative contribution of scattered, and hence more polarized, photons. At the bluest edge of the absorption, direct light is almost completely removed, and the residual flux may be dominated by scattered light, which would naturally enhance the observed polarization signal despite the low total intensity.

\subsection{Temporal Evolution of Line Effects}

\begin{figure}
\centering
\includegraphics[width=1.0\linewidth]{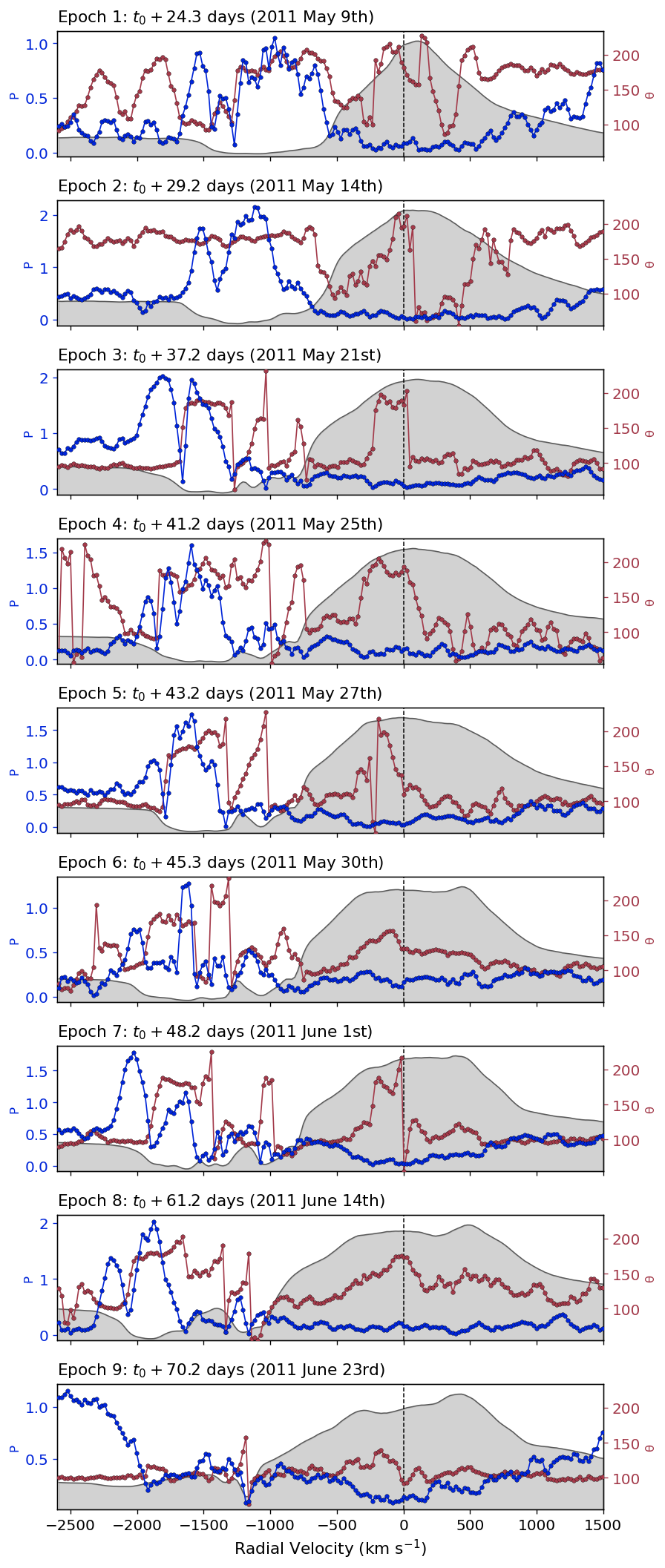}
\caption{Spectropolarimetric behaviour across the H$_\beta$ line for all observing epochs. Each panel corresponds to a different epoch of the 2011 eruption of T~Pyxidis. The intrinsic polarization degree $P_{\rm int}$ (blue; left axis) and polarization angle $\theta_{\rm int}$ (red; right axis) are shown as a function of radial velocity across the line profile. The grey shaded area represents the normalized flux profile of the H$_\beta$ P~Cygni line, and the vertical dashed line marks the rest velocity of the transition. This representation highlights the temporal evolution of the line polarization and the associated changes in the scattering geometry of the ejecta.}
\label{fig:zoomstacked}%
\end{figure}

The temporal evolution of the H$_\beta$ spectropolarimetric profile (Fig.~\ref{fig:zoomstacked}) reveals a progressive reorganization of the ejecta geometry during the eruption. Throughout the observing campaign, the polarization maxima remain systematically associated with the absorption troughs of the P-Cygni profile, while two preferred polarization orientations are observed, referred to as PA1 ($\approx 95^\circ$) and PA2 ($\approx 185^\circ$).

During the first three epochs, the line effect exhibits a clear bimodal structure. The fastest absorption component is generally associated with a residual polarization oriented along PA1, whereas the slower absorption system is generally associated with a residual polarization oriented along PA2. However, around optical maximum (Epoch 2), the line becomes almost entirely dominated by PA2, making this epoch unique within the observing sequence. By Epoch 3, the dual-axis configuration is recovered, and the first discrete absorption components (DACs) become apparent at intermediate velocities.

From Epochs 4 to 7, the spectropolarimetric evolution follows the spectroscopic development of the ejecta. As the emission profile broadens and develops multiple peaks, the overall spectropolarimetric morphology remains remarkably coherent, while the relative contribution of the two preferred polarization orientations evolves with time. The fastest absorption component remains preferentially aligned with PA1, while the slower absorption continues to trace PA2. During this phase, the DACs become increasingly prominent and display gradual rotations between the two preferred polarization orientations, suggesting that the relative contribution of the two dominant scattering geometries evolves with time.

The final observations indicate a simplification of the polarization structure. Although the dual-axis configuration is still visible at Epoch 8, the last epoch is almost entirely dominated by PA1, while the absorption components weaken and only a few DACs remain detectable. This evolution suggests that the ejecta progressively approach a single dominant scattering geometry as the nova enters its late decline.

Overall, the evolution of the line effects demonstrates that the polarization is closely linked to the kinematic structure responsible for the P-Cygni absorptions. The persistence of two preferred polarization orientations throughout most of the eruption indicates that at least two large-scale asymmetric components contribute to the scattering geometry, while their relative importance evolves as the ejecta expand.

\subsection{Comparison of the three Balmer lines}

\begin{figure*}[!t]
    \centering
    \includegraphics[width=\textwidth]{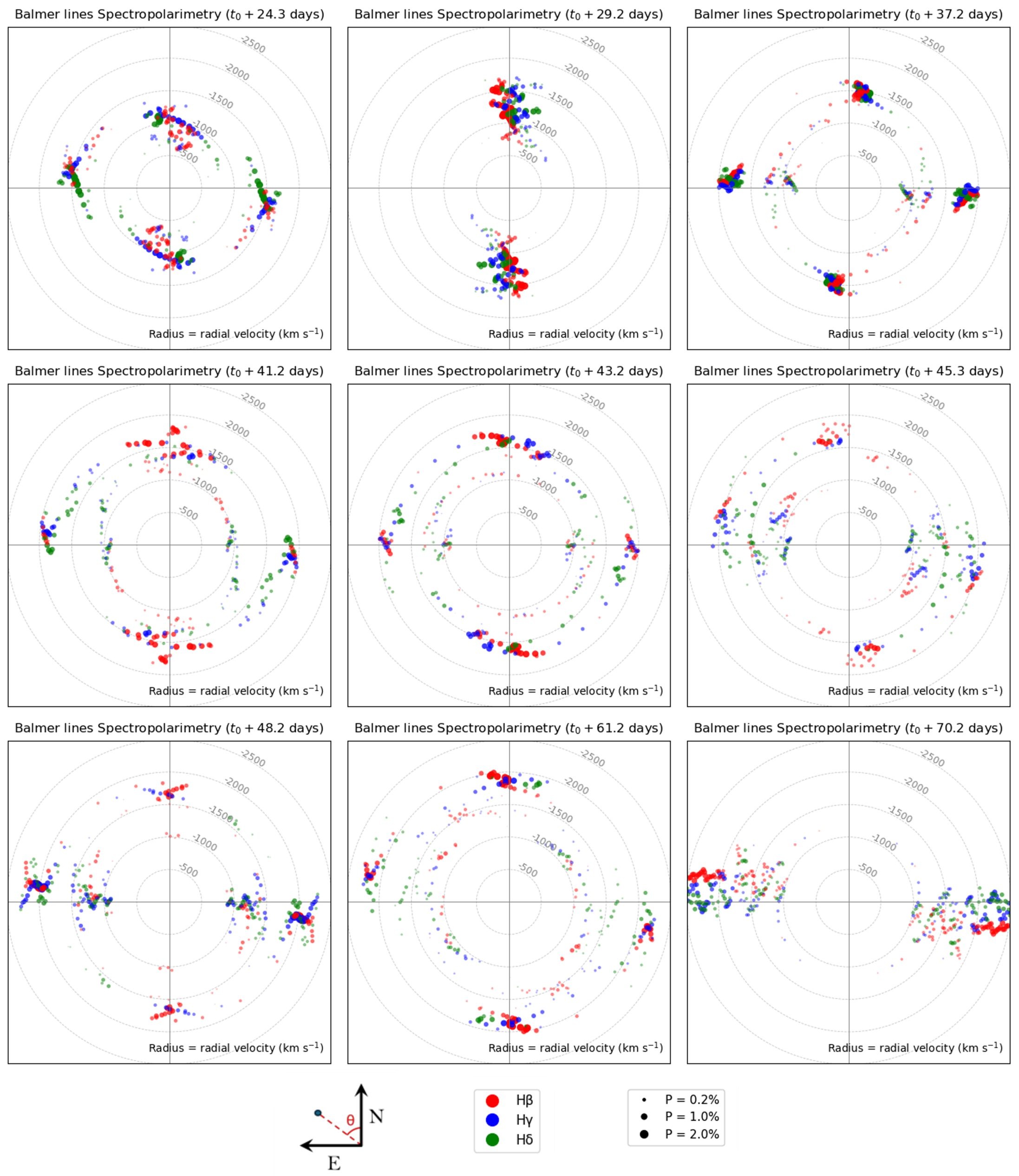}
    \caption{Representation of the spectropolarimetric properties of the Balmer lines (H$\beta$, H$\gamma$, H$\delta$) for each observing epoch. The radial velocity derived from the P~Cygni profiles is plotted as the distance from the origin, the polarization position angle defines the direction, and the degree of polarization is encoded in the point size. Colors identify the three Balmer transitions.}
\label{fig:new tous}
\end{figure*}

All Balmer lines originate in the hydrogen recombination region (inside the ionized sphere), but each probes a slightly different layer in depth or temperature. The H$\alpha$ line is usually the most intense, most extended, and most external, while the higher-order lines tend to be weaker, often arising from somewhat deeper, denser, and hotter layers. 

To compare the line effects across the three Balmer transitions, a graphical representation is introduced in which the degree and angle of the polarization are displayed together with the projected radial velocity (Fig.~\ref{fig:new tous}). For each pair of Stokes parameters (corrected for ISP and contamination), a point is plotted in an orthogonal system aligned with the sky orientation: 

\begin{itemize}
\item the degree of polarization  defines the size of the point,  
\item the polarization angle defines the angle between the point-origin vector and the north-south axis,  
\item the velocity associated with the point (derived from the P Cygni profile analysis) defines the distance of the point from the origin.  
\end{itemize}

This representation is conceptual and does not attempt to reflect the physical structure of the phenomenon. However, plotting the velocity as the distance from the origin is justified because the expansion velocity of the ejecta can be used as a proxy for their radius. 
This representation allows us to visualize the coupled evolution of velocity, polarization degree, and polarization angle across the Balmer lines. The points cluster at specific velocities and position angles rather than forming a continuous distribution, suggesting that the polarization maxima arise from distinct kinematic components within the ejecta. The fact that similar structures are observed simultaneously in H$_\beta$, H$_\gamma$, and H$_\delta$ indicates that these components affect multiple hydrogen transitions and likely reflect large-scale asymmetries in the expanding ejecta. The temporal evolution of these clusters reveals how the dominant polarizing structures evolve in velocity space as the eruption progresses.

\subsection{Evolution of the polarimetric peak velocities}

\begin{figure}
\centering
\includegraphics[width=\columnwidth]{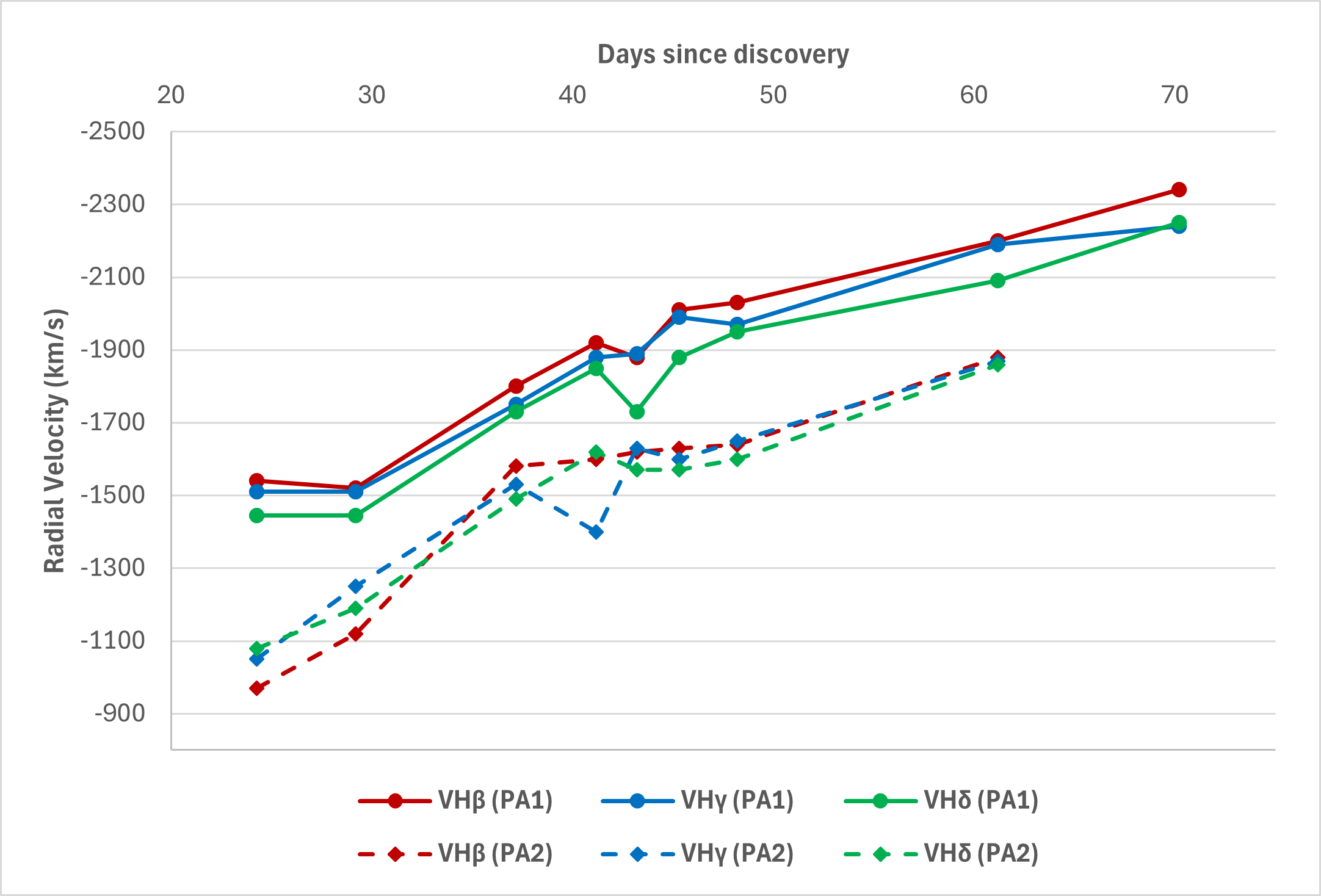}
\caption{Radial velocities of the polarization peaks associated with the absorption troughs of the Balmer lines (H$_\beta$, H$_\gamma$, H$_\delta$) as a function of time since discovery. Measurements are shown for the two dominant polarization position angles (PA $\approx$ 95° and PA $\approx$ 185°). The velocities increase with time, indicating that the polarimetric signal traces the evolving absorption system in the expanding ejecta.}
\label{fig:velocity}
\end{figure}

The polarization peaks associated with the absorption troughs of the Balmer lines show a clear systematic evolution with time. For each epoch we measured the radial velocity of the polarization maxima in H$_\beta$, H$_\gamma$, and H$_\delta$ for the two dominant polarization position angles (PA1 $\approx 95^{\circ}$ and PA2 $\approx 185^{\circ}$). The resulting velocities are shown in Fig.~\ref{fig:velocity} as functions of time. The figure displays the temporal evolution of the radial velocities associated with the polarization maxima in the Balmer lines. A clear trend is observed: in all three Balmer transitions, the velocity of the polarimetric peaks progressively shifts toward larger blueshifts as the eruption evolves. The effect is observed at both polarization position angles and follows a similar trend for the three lines. Because the polarization enhancement occurs at the velocity of the absorption troughs, the measured polarimetric peak velocities trace the kinematic evolution of the absorbing system. During the time interval covered by our observations, the velocity evolution shows an approximately linear increase over time.

This behavior closely mirrors the spectroscopic evolution reported for T\,Pyx during the 2011 eruption. High-resolution spectroscopy revealed that the absorption components of the Balmer lines progressively shift toward larger blueshifted velocities as the eruption evolves (Shore et al. 2011). These discrete absorption components observed in the P\,Cygni troughs were interpreted as signatures of a ballistic expansion of the ejecta, where progressively faster layers become observable as the optical depth decreases.

In this framework, the evolution of the absorption systems is governed by the propagation of the ionization/recombination front through the expanding envelope. As the ejecta expand and their density decreases, the ionized region progressively extends outward. Because nova ejecta follow an approximately ballistic velocity field ($v \propto r$), this outward propagation corresponds observationally to absorption systems appearing at progressively higher blueshifted velocities. As the ionization front reaches progressively larger radii in the expanding envelope, the absorbing structures responsible for the P\,Cygni profiles (and hence for the polarization peaks) are observed at increasingly higher expansion velocities.

The spectropolarimetric behavior reported here appears to follow the same evolution. The polarization peaks systematically coincide with the absorption troughs of the Balmer lines and therefore trace the same absorbing structures responsible for the P\,Cygni profiles. The quasi-linear increase of their radial velocities with time therefore suggests that the polarimetric signal follows the outward progression of the ionization/recombination front through the ballistic ejecta. The similar quasi-linear velocity evolution observed in H$_\beta$, H$_\gamma$, and H$_\delta$ suggests that the polarimetric peaks do not arise from line-specific effects, but rather trace a common physical front propagating through the expanding ejecta. This behavior is consistent with a globally ordered kinematic structure that undergoes homologous expansion.

\section{Discussion}

\subsection{Spectroscopic context}

The spectral evolution of T Pyx during the 2011 eruption has been extensively discussed in previous works, which showed that the nova evolved from an early He/N-like spectrum shortly after eruption to a 
\ion{Fe}{II}-dominated
phase around optical maximum, before transitioning again toward a later He/N phase during the decline. This behaviour was identified as evidence that the traditional “\ion{Fe}{II}” 
and “He/N” nova classes may not represent distinct categories, but rather different spectroscopic phases during the evolution of the ejecta \citep{Izzo+2012,Ederoclite2014}. 
More recently, \citet{Aydi+2024} proposed that this sequence reflects the progressive evolution of the ejecta opacity, ionization state, and density structure. They report that for T Pyx the 
\ion{Fe}{II} phase extends approximately from day 15 to day 40, while the late He/N phase begins around day 40 and continues until the nebular stage around day 150. 
The spectropolarimetric observations presented here cover the evolution of T Pyx from shortly before optical maximum through the early decline phase of the 2011 eruption. Thus, the polarimetric evolution observed here occurs precisely during the transition when deeper and/or geometrically distinct regions of the ejecta may begin to contribute to the observed spectrum.
Throughout most of the campaign, the optical spectra display strong Balmer P Cygni profiles together with prominent blueshifted absorption features characteristic of optically thick nova ejecta. The absorption components become barely detectable at the last epoch, indicating that the ejecta are evolving toward a more optically thin configuration. This behaviour is consistent with previous spectroscopic studies of the eruption \citep{Shore+2011,Surina+2014}. 
In this context, the spectropolarimetric evolution of T Pyx is likely tracing changes not only in the global geometry of the ejecta, but also in the relative contribution of different scattering and absorbing regions as the eruption evolves. The coexistence of strong line absorptions and evolving ionization conditions during the observed phases therefore provides a particularly favourable framework for investigating the three-dimensional structure of the ejecta through linear spectropolarimetry.

\subsection{Reliability of the intrinsic spectropolarimetric signal}

Interpreting the spectropolarimetric behaviour of novae critically depends on the accurate determination of the interstellar polarization (ISP). While the ISP contribution mainly acts as a translation in the Q-U plane and therefore does not modify the morphology of intrinsic structures such as loops or line excursions, it can strongly affect the polarization degree and, more importantly, the inferred position angle because of the non-linear transformation between the Stokes parameters and $(P,\theta)$. An incorrect ISP estimate may therefore artificially introduce apparent rotations of the polarization angle, modify the relative orientation of different structures, or even mask intrinsic symmetries in the ejecta. Since the interpretation developed in this work relies primarily on the evolution of the polarization angle and on the existence of preferential polarization axes, establishing the reliability of the ISP correction is essential.
The ISP determination adopted here is based on the depolarization observed at the rest wavelength of the Balmer recombination lines, under the assumption that the line emission itself is intrinsically unpolarized and dilutes the polarized continuum. This approach is commonly used in nova spectropolarimetry  as line photons typically form over more extended regions and experience fewer electron scatterings, leading to a dilution of the polarized continuum \citep[e.g.][]{McLean1979,Bjorkman+1994,Maund+2007}.
In this work, this assumption is further supported by the temporal stability of the polarization signal measured within the Balmer emission components: The observed Stokes parameters around the rest wavelengths of the Balmer recombination lines show minimal dispersion within a narrow spectral interval and remain remarkably stable across the different epochs. Such behaviour is naturally expected for an interstellar contribution, whereas intrinsic ejecta polarization would be expected to evolve significantly during the strong spectroscopic changes observed throughout the eruption. It strongly suggests that the polarization measured at the rest wavelengths of the hydrogen recombination lines is not significantly affected by intrinsic variability in the ejecta. 
An additional argument supporting the robustness of the ISP determination is the consistency obtained independently from the three Balmer lines. The polarization values derived at the rest wavelengths of H$_\beta$, H$_\gamma$, and H$_\delta$ are mutually compatible and define a coherent wavelength dependence well reproduced by a Serkowski law and can therefore be used as a reliable approximation for the ISP.

\subsection{Continuum polarization}

The continuum polarization of T Pyx reveals the presence of a persistent intrinsic asymmetry throughout most of the observing campaign. After subtraction of the interstellar contribution, the continuum polarization remains typically below $1$\%, but displays a remarkably stable polarization position angle clustered around $PA1 \approx 95^\circ$. Such stability strongly suggests that the continuum-forming region departs significantly from spherical symmetry and possesses a long-lived axisymmetric geometry projected on the plane of the sky. The fact that this dominant polarization orientation is already present before optical maximum and persists during the decline further indicates that the large-scale ejecta geometry is established at very early stages of the eruption and remains globally stable over time.
A notable exception occurs at Epoch 2, near optical maximum, when the continuum polarization angle undergoes an abrupt rotation of approximately 90°, becoming aligned close to $PA2 \approx 185^\circ$. Unlike the other epochs, during which the continuum polarization is predominantly oriented along PA1, the spectropolarimetric signal at maximum light becomes almost entirely dominated by this orthogonal orientation. Such a sudden rotation is unlikely to result from uncertainties in the ISP correction or from stochastic fluctuations in the ejecta. Instead, it strongly suggests that the relative contribution of different asymmetric regions changes significantly around maximum light.
Importantly, a rotation of the observed polarization angle does not necessarily imply a physical reorientation of the ejecta themselves. In electron-scattering dominated envelopes, the observed polarization depends not only on the intrinsic geometry of the system, but also on the relative visibility and weighting of the different scattering regions. Changes in optical depth may therefore modify which regions dominate the observed polarized flux, producing apparent rotations of the continuum polarization angle without requiring any global structural rearrangement of the ejecta \citep[e.g.][]{Hoflich+1991,Schulte-Ladbeck+1992}. 
This interpretation is naturally consistent with the expected evolution of the nova pseudo-photosphere. Around optical maximum, the ejecta are highly optically thick and the pseudo-photosphere expands to very large radii, temporarily obscuring the innermost regions of the system \citep{Chomiuk+2021}. 
As the eruption evolves and the ejecta progressively become more transparent, the effective continuum-forming region recedes inward and deeper structures begin to contribute to the observed polarized flux. The transient dominance of the PA2 component near maximum light may therefore reflect a temporary change in the relative visibility of distinct asymmetric regions as the optical depth evolves, rather than the emergence of a fundamentally new geometry.
Such behaviour is qualitatively similar to that observed in several asymmetric eruptive systems, including Nova Cyg 1992 \citep{Bjorkman+1994}, 
HD 45677 \citep{Schulte-Ladbeck+1992}, 
SN 2009ip \citep{Mauerhan+2014}, 
V339 Del \citep{Kawakita+2019}, 
and RS Oph \citep{Nikolov+2023}, 
where orthogonal polarization components have been interpreted as the signature of multiple asymmetric scattering regions whose relative contribution evolves with time. In the case of T Pyx, the persistence of the PA1 component throughout most of the eruption suggests that it traces the dominant large-scale asymmetry of the continuum-forming region, while the temporary appearance of the orthogonal PA2 component near optical maximum likely reflects evolving optical-depth effects within a more complex ejecta morphology.
The continuum spectropolarimetry therefore indicates that the ejecta of T Pyx cannot be described by a simple homogeneous spherical outflow. Instead, the observations strongly support the presence of a structured asymmetric continuum-forming region whose observed polarization properties evolve as the eruption expands and progressively reveals deeper layers of the ejecta.

\subsection{Polarization across the Balmer absorptions}

The continuum analysis suggests that more than one asymmetric scattering region contributes to the observed polarization. If several polarized continuum components coexist, absorption features are expected to modify their relative contributions by selectively occulting different regions of the photosphere. 
The spectropolarimetric signatures observed across the Balmer lines reveal a substantially more complex ejecta geometry than inferred from the continuum polarization alone. Strong intrinsic polarization peaks are systematically detected across the blueshifted absorption troughs of the H$_\beta$, H$_\gamma$, and H$_\delta$ P-Cygni profiles, frequently reaching values of $1\text{--}2\,\%$. These line effects are accompanied by clear excursions and loops in the Q-U plane, demonstrating that the polarization angle varies across the line profile and therefore that the line-forming regions cannot be described by a simple axisymmetric geometry. The remarkable consistency of these signatures among the three Balmer transitions strongly indicates that they trace large-scale structures affecting the global hydrogen recombination region rather than isolated line-specific phenomena or blending effects. 
The close association between the polarization peaks and the blueshifted absorption components naturally supports the selective obscuration scenario originally proposed by \citet{McLean1979}. In this framework, the line polarization does not primarily arise from intrinsically polarized line emission, but instead results from asymmetric absorbing material selectively occulting different regions of an already polarized continuum-forming photosphere. Because the absorbing ejecta remove flux anisotropically across the projected surface of the source, the cancellation of the continuum polarization becomes incomplete, producing enhanced polarization within the absorption troughs. The disappearance of the polarization signatures at the final epoch, simultaneously with the weakening of the Balmer absorptions in the flux spectra, strongly supports this interpretation and directly links the observed line polarization to the absorbing structures responsible for the P-Cygni profiles. In this picture, the polarization measured across an absorption component does not directly trace the geometry of the absorbing material itself. Instead, it reflects the polarized continuum that remains visible after selective occultation. The Balmer absorptions therefore act as natural masks that isolate the contribution of different continuum-scattering regions.
A particularly remarkable characteristic of the Balmer polarization profiles is the recurrent presence of multiple polarization maxima within a single absorption trough. In most epochs, the absorption-induced polarization is composed of at least two dominant components associated with distinct polarization orientations, frequently close to orthogonal. These components are generally associated with different velocity intervals within the absorption profile (with the higher-velocity absorptions preferentially associated with PA1, while the lower-velocity absorptions tend to be associated with PA2) and are interpreted as the polarized continuum that remains after selective occultation by absorbing material at those velocities. Such behaviour indicates that different velocity layers of the ejecta contribute differently to the observed polarization and strongly suggests that the Balmer-forming region possesses a multi-component geometry rather than a single homogeneous asymmetric structure. If the continuum polarization arises from the superposition of several asymmetric scattering regions, an absorption component preferentially associated with one of these regions will suppress its contribution to the observed polarized flux. The resulting polarization angle therefore becomes dominated by the complementary scattering component.
Interestingly, the high-velocity polarization component is generally associated with a relatively shallow absorption feature, whereas the strongest absorption troughs are typically dominated by the lower-velocity component. This behaviour suggests that the high-velocity material covers a smaller fraction of the projected continuum-forming region than the slower absorbing ejecta. Despite its weaker imprint in the flux spectrum, the high-velocity component produces a strong polarization signal, indicating that its projected geometry is highly asymmetric and therefore particularly efficient at breaking the cancellation of the continuum polarization. In contrast, the lower-velocity component appears to occult a larger fraction of the photosphere, producing deeper absorption while preserving a more symmetric distribution of the polarized flux. Such behaviour is naturally expected in partial obscuration scenarios, where the observed polarization depends primarily on the asymmetry and covering factor of the absorbing material rather than solely on its absorption strength.
Importantly, the existence of multiple polarization components is considerably clearer in the spectropolarimetric data than in the flux spectra themselves. While the absorption troughs often appear relatively smooth in intensity, the polarization profiles reveal distinct geometrical contributions that remain partially blended in flux space. The spectropolarimetry therefore provides access not only to the kinematics of the ejecta, but also to their projected geometrical structure. In this sense, the velocity-resolved polarization profiles act as a tomographic probe of the expanding ejecta, with different Doppler velocities sampling different asymmetric regions along the line of sight. 
The loops observed in the Q-U plane further support this interpretation. In a purely axisymmetric configuration characterized by a single dominant polarization orientation, the polarization across a spectral line is expected to follow approximately linear excursions in Q-U space. In contrast, the presence of loops implies that both the degree and orientation of polarization vary continuously across the line profile, revealing departures from simple axial symmetry and indicating that different velocity intervals probe different projected geometries. The recurrent detection of such loops throughout the eruption therefore strongly supports the presence of multiple asymmetric regions simultaneously contributing to the observed polarization signal. 
Despite their different polarization orientations, the various absorption components display a remarkably coherent temporal evolution. In all three Balmer lines, the velocities associated with the polarization maxima progressively shift toward larger blueshifts with time, following similar quasi-linear trends for the different polarization orientations. This behaviour suggests that the different polarimetric components are not entirely independent ejecta episodes evolving separately, but more likely correspond to geometrically distinct regions embedded within the same global expanding outflow. The observations therefore support a structured ejecta morphology in which multiple asymmetric regions coexist while sharing a common large-scale dynamical evolution. 
The spectropolarimetric behaviour observed near optical maximum is particularly noteworthy. At Epoch 2, both the continuum polarization and the Balmer absorption polarization become almost entirely aligned with PA2, whereas the orthogonal PA1 component largely disappears. This behaviour mirrors the evolution observed in the continuum polarization and suggests that the relative visibility of the different asymmetric regions changes significantly around maximum light. As discussed previously, such behaviour may naturally arise from optical-depth effects within an evolving pseudo-photosphere, temporarily modifying the relative contribution of the various scattering and absorbing regions without requiring a fundamental reorganization of the ejecta geometry itself. 
Overall, the polarization behaviour across the Balmer absorptions demonstrates that the ejecta of T Pyx possess a complex multi-component geometry whose projected structure varies with both velocity and time. The observations strongly indicate that different regions of the expanding ejecta contribute differently to the polarized flux and that the line-forming region cannot be described by a single global axisymmetric structure.

\subsection{Towards a geometrical interpretation}

The spectropolarimetric observations presented here demonstrate that the ejecta of T Pyx possess a complex and evolving intrinsic geometry already during the earliest optically thick phases of the eruption. The continuum polarization reveals the presence of a persistent large-scale asymmetry characterized by a dominant polarization orientation ($PA1 \approx 95^\circ$) that remains remarkably stable throughout most of the observing campaign. Superimposed on this global structure, the Balmer absorptions display strong velocity-dependent polarization signatures associated with multiple polarization orientations, frequently close to orthogonal. The coexistence of these components, together with the recurrent loops observed in the Q-U plane, indicates that the ejecta cannot be described by a single homogeneous axisymmetric outflow. Instead, the observations strongly support a structured multi-component morphology whose projected geometry varies with both velocity and time.
The spectropolarimetric behaviour of the Balmer absorptions further suggests that different velocity layers probe distinct asymmetric regions of the ejecta. In particular, the recurrent separation between a weakly absorbing high-velocity polarization component and a strongly absorbing lower-velocity component indicates significant differences in covering factor and projected geometry within the expanding outflow. The weaker absorption depth associated with the high-velocity component suggests that this material occults only a limited fraction of the continuum-forming region, while its strong polarization signature implies a highly asymmetric projected distribution that efficiently breaks the cancellation of the continuum polarization. Conversely, the lower-velocity absorptions appear to occult a larger fraction of the photosphere and therefore dominate the absorption troughs while producing a different net polarization orientation. Such behaviour is naturally expected in partial obscuration scenarios, where the observed polarization depends primarily on the asymmetry and covering factor of the absorbing structures rather than solely on their absorption strength.
Within this framework, the polarization angle observed across a given absorption component does not necessarily trace the geometry of the absorbing material itself, but rather the dominant polarization orientation of the continuum regions that remain visible after selective occultation. A localized absorbing structure may therefore suppress one polarized component and enhance the relative contribution of another. The coexistence of the two dominant polarization orientations observed in T Pyx may consequently reflect the simultaneous contribution of multiple large-scale asymmetric regions within the ejecta.
Under standard electron-scattering geometries, the observed polarization vector is expected to be perpendicular to the dominant projected symmetry axis of the scattering structure. The persistent continuum polarization angle $PA1 \approx 95^\circ$ therefore implies a projected structural axis close to 5°. Remarkably, this orientation is in excellent agreement with the morphology independently inferred from recent spatially resolved observations of T Pyx. Using MUSE observations of the nebular remnant, \citet{Santamaria+2025} 
identified an ellipsoidal and toroidal H$_\beta$-emitting structure oriented along $PA \approx 5^\circ$, together with bipolar ejecta components approximately orthogonal to this direction. Likewise, near-infrared interferometric observations obtained during the 2011 eruption by \citet{Chesneau+2011} 
were successfully reproduced using a nearly face-on bipolar outflow model with a polar-axis orientation of $PA \approx 110^\circ$. Although the exact position angles are not strictly identical, the overall agreement between the spectropolarimetric and spatially resolved geometries is striking.
In this context, the dominant continuum polarization component observed at PA1 may naturally be associated with scattering within an equatorial or toroidal structure whose projected major axis lies close to north-south, while the orthogonal PA2 component may reflect the contribution of material distributed preferentially along the bipolar direction. Within the selective obscuration framework, absorption occurring predominantly within the bipolar component would preferentially suppress the corresponding polarized flux and therefore enhance the relative contribution of the equatorial polarization component, producing line polarization aligned with PA1. Conversely, stronger absorption arising within the equatorial structure would preferentially attenuate the equatorial polarized flux, leaving the bipolar component dominant and producing polarization aligned with PA2. Such a scenario qualitatively accounts for the recurrent association of the weak high-velocity absorption component with PA1, and the stronger low-velocity absorption component with PA2.
The abrupt polarization rotation observed near optical maximum likely results from changes in the relative visibility of these asymmetric regions as the optical depth of the ejecta evolves. Around maximum light, the highly extended pseudo-photosphere may temporarily obscure some inner scattering regions, modifying the balance between the competing polarization components. As the ejecta expand and progressively become optically thinner, deeper structures re-emerge and the dominant continuum polarization returns to PA1. The spectropolarimetric evolution therefore appears to be governed not by a physical reorientation of the ejecta, but by the changing contribution of different asymmetric regions during the recession of the continuum-forming pseudo-photosphere.
Although the exact three-dimensional morphology of the ejecta cannot be uniquely constrained from the spectropolarimetric data alone, the remarkable consistency between the observed polarization orientations and the independently resolved structures strongly suggests that the large-scale asymmetries observed in the mature nebular remnant are already established during the earliest stages of the eruption. The observations presented here therefore demonstrate the diagnostic power of linear spectropolarimetry for probing the three-dimensional geometry of nova ejecta during phases that remain spatially unresolved, and indicate that the ejecta of T Pyx are intrinsically structured, multi-component, and significantly asymmetric from the onset of the eruption.
The spectropolarimetric results presented here therefore bridge the gap between the early unresolved phases of the eruption and the later spatially resolved morphology of the remnant, suggesting that the equatorial and bipolar structures identified at late times were already established and detectable through optical polarization during the first months after outburst.

\section{Conclusions}

%

We have presented the first multi-epoch optical linear spectropolarimetric study of the 2011 eruption of the recurrent nova T Pyx, covering the phases from shortly before optical maximum to the early decline. After careful correction for instrumental and interstellar polarization, the observations reveal clear intrinsic continuum and line polarization signatures demonstrating significant departures from spherical symmetry from the earliest optically thick stages of the eruption.
The continuum polarization is characterized by a dominant and persistent polarization orientation ($PA1\approx95^\circ$), indicating the presence of a stable large-scale asymmetric structure throughout most of the observing campaign. A temporary rotation toward an orthogonal polarization component ($PA2\approx185^\circ$) near optical maximum is interpreted as the consequence of evolving optical-depth effects within the expanding pseudo-photosphere rather than a physical reorientation of the ejecta.
Strong polarization peaks associated with the Balmer P-Cygni absorptions reveal a substantially more complex geometry than inferred from the continuum alone. The velocity-dependent polarization signatures, recurrent orthogonal polarization components, and loops observed in the Q-U plane demonstrate that the ejecta possess a structured multi-component morphology whose projected geometry varies with both velocity and time. The spectropolarimetric behaviour is naturally explained within a selective obscuration framework involving multiple asymmetric scattering and absorbing regions with different covering factors.
Assuming standard electron-scattering geometries, the dominant continuum polarization angle implies a projected structural axis close to $PA \approx 5^\circ$, remarkably consistent with the equatorial/toroidal morphology independently inferred from recent interferometric and MUSE observations of the T Pyx remnant. The orthogonal polarization component may similarly trace the bipolar ejecta component resolved at later stages. The spectropolarimetric results presented here therefore bridge the gap between the early unresolved phases of the eruption and the later spatially resolved morphology of the remnant, suggesting that the large-scale equatorial and bipolar structures are already established during the first months after outburst.
These observations illustrate the unique diagnostic power of linear spectropolarimetry for probing the three-dimensional structure of nova ejecta during phases that remain spatially unresolved and highlight the importance of time-resolved spectropolarimetric monitoring for understanding the origin and evolution of asymmetries in nova explosions.

\begin{acknowledgements}
AE acknowledges the financial support from the Spanish Ministry of Science and Innovation and the European Union - NextGenerationEU through the Recovery and Resilience Facility project ICTS-MRR-2021-03-CEFCA.
This work is based on observations collected at the European Southern Observatory under ESO program 287.D-5024(A).
The authors have used ChatGPT (OpenAI; model version 5.6) as support for developing and refining the Python analysis and visualization software, and for language editing of the manuscript. All AI-assisted outputs were reviewed and verified by the authors, who take full responsibility for the content of the publication.
\end{acknowledgements}

%
%


\bibliographystyle{aa} 
\bibliography{tpyx} 

@ARTICLE{Aydi+2020,
       author = {{Aydi}, E. and {Chomiuk}, L. and {Izzo}, L. and {Harvey}, E.~J. and {Leahy-McGregor}, J. and {Strader}, J. and {Buckley}, D.~A.~H. and {Sokolovsky}, K.~V. and {Kawash}, A. and {Kochanek}, C.~S. and {Linford}, J.~D. and {Metzger}, B.~D. and {Mukai}, K. and {Orio}, M. and {Shappee}, B.~J. and {Shishkovsky}, L. and {Steinberg}, E. and {Swihart}, S.~J. and {Sokoloski}, J.~L. and {Walter}, F.~M. and {Woudt}, P.~A.},
        title = "{Early Spectral Evolution of Classical Novae: Consistent Evidence for Multiple Distinct Outflows}",
      journal = {\apj},
         year = 2020,
        month = dec,
       volume = {905},
       number = {1},
          eid = {62},
        pages = {62},
          doi = {10.3847/1538-4357/abc3bb},
archivePrefix = {arXiv},
       eprint = {2010.07481},
 primaryClass = {astro-ph.HE},
       adsurl = {https://ui.adsabs.harvard.edu/abs/2020ApJ...905...62A}
}

@ARTICLE{Aydi+2024,
       author = {{Aydi}, E. and {Chomiuk}, L. and {Strader}, J. and {Sokolovsky}, K.~V. and {Williams}, R.~E. and {Buckley}, D.~A.~H. and {Ederoclite}, A. and {Izzo}, L. and {Kyer}, R. and {Linford}, J.~D. and {Kniazev}, A. and {Metzger}, B.~D. and {Miko{\l}ajewska}, J. and {Molaro}, P. and {Molina}, I. and {Mukai}, K. and {Munari}, U. and {Orio}, M. and {Panurach}, T. and {Shappee}, B.~J. and {Shen}, K.~J. and {Sokoloski}, J.~L. and {Urquhart}, R. and {Walter}, F.~M.},
        title = "{Revisiting the classics: on the evolutionary origin of the 'Fe II' and 'He/N' spectral classes of novae}",
      journal = {\mnras},
         year = 2024,
        month = jan,
       volume = {527},
       number = {3},
        pages = {9303-9321},
          doi = {10.1093/mnras/stad3342},
archivePrefix = {arXiv},
       eprint = {2309.07097},
 primaryClass = {astro-ph.SR},
       adsurl = {https://ui.adsabs.harvard.edu/abs/2024MNRAS.527.9303A}
}

@ARTICLE{Bjorkman+1994,
       author = {{Bjorkman}, K.~S. and {Johansen}, K.~A. and {Nordsieck}, K.~H. and {Gallagher}, J.~S. and {Barger}, A.~J.},
        title = "{Spectropolarimetry of Nova Cygni 1992: Evidence for an Asymmetric Geometry}",
      journal = {\apj},
         year = 1994,
        month = apr,
       volume = {425},
        pages = {247},
          doi = {10.1086/173981},
       adsurl = {https://ui.adsabs.harvard.edu/abs/1994ApJ...425..247B}
}

@BOOK{Bode+Evans,
       author = {{Bode}, Michael F. and {Evans}, Aneurin},
        title = "{Classical Novae}",
         year = 2008,
       volume = {43},
          doi = {10.1017/CBO9780511536168},
       adsurl = {https://ui.adsabs.harvard.edu/abs/2008clno.book.....B}
}

@ARTICLE{Chesneau+2011,
       author = {{Chesneau}, O. and {Meilland}, A. and {Banerjee}, D.~P.~K. and {Le Bouquin}, J.-B. and {McAlister}, H. and {Millour}, F. and {Ridgway}, S.~T. and {Spang}, A. and {ten Brummelaar}, T. and {Wittkowski}, M. and {Ashok}, N.~M. and {Benisty}, M. and {Berger}, J.-P. and {Boyajian}, T. and {Farrington}, Ch. and {Goldfinger}, P.~J. and {Merand}, A. and {Nardetto}, N. and {Petrov}, R. and {Rivinius}, Th. and {Schaefer}, G. and {Touhami}, Y. and {Zins}, G.},
        title = "{The 2011 outburst of the recurrent nova <ASTROBJ>T Pyxidis</ASTROBJ>. Evidence for a face-on bipolar ejection}",
      journal = {\aap},
         year = 2011,
        month = oct,
       volume = {534},
          eid = {L11},
        pages = {L11},
          doi = {10.1051/0004-6361/201117792},
archivePrefix = {arXiv},
       eprint = {1109.4534},
 primaryClass = {astro-ph.SR},
       adsurl = {https://ui.adsabs.harvard.edu/abs/2011A&A...534L..11C}
}

@ARTICLE{Chomiuk+2014Natur,
       author = {{Chomiuk}, Laura and {Linford}, Justin D. and {Yang}, Jun and {O'Brien}, T.~J. and {Paragi}, Zsolt and {Mioduszewski}, Amy J. and {Beswick}, R.~J. and {Cheung}, C.~C. and {Mukai}, Koji and {Nelson}, Thomas and {Ribeiro}, Val{\'e}rio A.~R.~M. and {Rupen}, Michael P. and {Sokoloski}, J.~L. and {Weston}, Jennifer and {Zheng}, Yong and {Bode}, Michael F. and {Eyres}, Stewart and {Roy}, Nirupam and {Taylor}, Gregory B.},
        title = "{Binary orbits as the driver of {\ensuremath{\gamma}}-ray emission and mass ejection in classical novae}",
      journal = {\nat},
         year = 2014,
        month = oct,
       volume = {514},
       number = {7522},
        pages = {339-342},
          doi = {10.1038/nature13773},
archivePrefix = {arXiv},
       eprint = {1410.3473},
 primaryClass = {astro-ph.HE},
       adsurl = {https://ui.adsabs.harvard.edu/abs/2014Natur.514..339C}
}

@ARTICLE{Chomiuk+2021,
       author = {{Chomiuk}, Laura and {Metzger}, Brian D. and {Shen}, Ken J.},
        title = "{New Insights into Classical Novae}",
      journal = {\araa},
         year = 2021,
        month = sep,
       volume = {59},
        pages = {391-444},
          doi = {10.1146/annurev-astro-112420-114502},
archivePrefix = {arXiv},
       eprint = {2011.08751},
 primaryClass = {astro-ph.HE},
       adsurl = {https://ui.adsabs.harvard.edu/abs/2021ARA&A..59..391C}
}

@ARTICLE{Cikota+2017,
       author = {{Cikota}, Aleksandar and {Patat}, Ferdinando and {Cikota}, Stefan and {Faran}, Tamar},
        title = "{Linear spectropolarimetry of polarimetric standard stars with VLT/FORS2}",
      journal = {\mnras},
         year = 2017,
        month = feb,
       volume = {464},
       number = {4},
        pages = {4146-4159},
          doi = {10.1093/mnras/stw2545},
archivePrefix = {arXiv},
       eprint = {1610.00722},
 primaryClass = {astro-ph.IM},
       adsurl = {https://ui.adsabs.harvard.edu/abs/2017MNRAS.464.4146C}
}

@ARTICLE{Csak+2005,
       author = {{Cs{\'a}k}, B. and {Kiss}, L.~L. and {Retter}, A. and {Jacob}, A. and {Kaspi}, S.},
        title = "{Spectroscopic monitoring of the transition phase in nova <ASTROBJ>V4745 Sgr</ASTROBJ>}",
      journal = {\aap},
         year = 2005,
        month = jan,
       volume = {429},
        pages = {599-605},
          doi = {10.1051/0004-6361:20035751},
archivePrefix = {arXiv},
       eprint = {astro-ph/0408268},
 primaryClass = {astro-ph},
       adsurl = {https://ui.adsabs.harvard.edu/abs/2005A&A...429..599C}
}

@INPROCEEDINGS{Ederoclite2014,
       author = {{Ederoclite}, A.},
        title = "{The Mystery of T Pyxidis, The 2011 Explosion}",
    booktitle = {Stellar Novae: Past and Future Decades},
         year = 2014,
       editor = {{Woudt}, P.~A. and {Ribeiro}, V.~A.~R.~M.},
       series = {Astronomical Society of the Pacific Conference Series},
       volume = {490},
        month = dec,
        pages = {163},
          doi = {10.48550/arXiv.1304.1305},
archivePrefix = {arXiv},
       eprint = {1304.1305},
 primaryClass = {astro-ph.SR},
       adsurl = {https://ui.adsabs.harvard.edu/abs/2014ASPC..490..163E}
}

@ARTICLE{Figueira+2018,
       author = {{Figueira}, Joana and {Jos{\'e}}, Jordi and {Garc{\'\i}a-Berro}, Enrique and {Campbell}, Simon W. and {Garc{\'\i}a-Senz}, Domingo and {Mohamed}, Shazrene},
        title = "{Three-dimensional simulations of the interaction between the nova ejecta, accretion disk, and companion star}",
      journal = {\aap},
         year = 2018,
        month = may,
       volume = {613},
          eid = {A8},
        pages = {A8},
          doi = {10.1051/0004-6361/201731545},
archivePrefix = {arXiv},
       eprint = {1712.08402},
 primaryClass = {astro-ph.SR},
       adsurl = {https://ui.adsabs.harvard.edu/abs/2018A&A...613A...8F}
}

@ARTICLE{Gallagher+1978,
       author = {{Gallagher}, J.~S. and {Starrfield}, S.},
        title = "{Theory and observations of classical novae.}",
      journal = {\araa},
         year = 1978,
        month = jan,
       volume = {16},
        pages = {171-214},
          doi = {10.1146/annurev.aa.16.090178.001131},
       adsurl = {https://ui.adsabs.harvard.edu/abs/1978ARA&A..16..171G}
}

@ARTICLE{Gilmozzi+Selvelli2007,
       author = {{Gilmozzi}, R. and {Selvelli}, P.},
        title = "{The secrets of T Pyxidis. I. UV observations}",
      journal = {\aap},
         year = 2007,
        month = jan,
       volume = {461},
       number = {2},
        pages = {593-603},
          doi = {10.1051/0004-6361:20054182},
archivePrefix = {arXiv},
       eprint = {astro-ph/0610028},
 primaryClass = {astro-ph},
       adsurl = {https://ui.adsabs.harvard.edu/abs/2007A&A...461..593G}
}

@ARTICLE{Hoflich+1991,
       author = {{Hoflich}, P.},
        title = "{Asphericity effects in scatterring dominated photospheres.}",
      journal = {\aap},
         year = 1991,
        month = jun,
       volume = {246},
        pages = {481},
       adsurl = {https://ui.adsabs.harvard.edu/abs/1991A&A...246..481H}
}

@ARTICLE{Hutchings1970,
       author = {{Hutchings}, J.~B.},
        title = "{A Spectrographic Analysis of Nova Vulpeculae 1968, No. 1}",
      journal = {\pasp},
         year = 1970,
        month = jun,
       volume = {82},
       number = {487},
        pages = {603},
          doi = {10.1086/128937},
       adsurl = {https://ui.adsabs.harvard.edu/abs/1970PASP...82..603H}
}

@ARTICLE{Izzo+2012,
       author = {{Izzo}, L. and {Ederoclite}, A. and {Della Valle}, M. and {Mason}, E. and {Williams}, R.~E. and {Altamore}, T. and {Cassatella}, A. and {Gilmozzi}, R. and {Patat}, F. and {Schmidtobreick}, L. and {Selvelli}, P. and {Tappert}, C. and {Thater}, S. and {Covone}, G. and {Dall'Ora}, M. and {Paolillo}, M.},
        title = "{Optical and near infrared multi-site follow up of the recurrent nova T Pyx}",
      journal = {\memsai},
         year = 2012,
        month = jan,
       volume = {83},
        pages = {830},
       adsurl = {https://ui.adsabs.harvard.edu/abs/2012MmSAI..83..830I}
}

@ARTICLE{Kawakita+2019,
       author = {{Kawakita}, H. and {Shinnaka}, Y. and {Arai}, A. and {Arasaki}, T. and {Ikeda}, Y.},
        title = "{High-resolution Optical Spectropolarimetry of Nova V339 Del: Spatial Distribution of Nova Ejecta during the Early Phase of Explosion}",
      journal = {\apj},
         year = 2019,
        month = feb,
       volume = {872},
       number = {2},
          eid = {120},
        pages = {120},
          doi = {10.3847/1538-4357/aaff68},
       adsurl = {https://ui.adsabs.harvard.edu/abs/2019ApJ...872..120K}
}

@ARTICLE{Mason+2018,
       author = {{Mason}, Elena and {Shore}, Steven N. and {De Gennaro Aquino}, Ivan and {Izzo}, Luca and {Page}, Kim and {Schwarz}, Greg J.},
        title = "{V1369 Cen High-resolution Panchromatic Late Nebular Spectra in the Context of a Unified Picture for Nova Ejecta}",
      journal = {\apj},
         year = 2018,
        month = jan,
       volume = {853},
       number = {1},
          eid = {27},
        pages = {27},
          doi = {10.3847/1538-4357/aaa247},
archivePrefix = {arXiv},
       eprint = {1807.07178},
 primaryClass = {astro-ph.SR},
       adsurl = {https://ui.adsabs.harvard.edu/abs/2018ApJ...853...27M}
}

@ARTICLE{Mauerhan+2014,
       author = {{Mauerhan}, Jon and {Williams}, G. Grant and {Smith}, Nathan and {Smith}, Paul S. and {Filippenko}, Alexei V. and {Hoffman}, Jennifer L. and {Milne}, Peter and {Leonard}, Douglas C. and {Clubb}, Kelsey I. and {Fox}, Ori D. and {Kelly}, Patrick L.},
        title = "{Multi-epoch spectropolarimetry of SN 2009ip: direct evidence for aspherical circumstellar material}",
      journal = {\mnras},
         year = 2014,
        month = aug,
       volume = {442},
       number = {2},
        pages = {1166-1180},
          doi = {10.1093/mnras/stu730},
archivePrefix = {arXiv},
       eprint = {1403.4240},
 primaryClass = {astro-ph.SR},
       adsurl = {https://ui.adsabs.harvard.edu/abs/2014MNRAS.442.1166M}
}

@ARTICLE{Maund+2007,
       author = {{Maund}, Justyn R. and {Wheeler}, J. Craig and {Patat}, Ferdinando and {Baade}, Dietrich and {Wang}, Lifan and {H{\"o}flich}, Peter},
        title = "{Spectropolarimetry of the Type Ib/c SN 2005bf}",
      journal = {\mnras},
         year = 2007,
        month = oct,
       volume = {381},
       number = {1},
        pages = {201-210},
          doi = {10.1111/j.1365-2966.2007.12230.x},
archivePrefix = {arXiv},
       eprint = {0707.2237},
 primaryClass = {astro-ph},
       adsurl = {https://ui.adsabs.harvard.edu/abs/2007MNRAS.381..201M}
}

@ARTICLE{McLaughlin1943,
       author = {{McLaughlin}, Dean Benjamin},
        title = "{On the spectra of novae}",
      journal = {Publications of Michigan Observatory},
         year = 1943,
        month = jan,
       volume = {8},
       number = {12},
        pages = {149-194},
       adsurl = {https://ui.adsabs.harvard.edu/abs/1943POMic...8..149M}
}

@ARTICLE{McLean1979,
       author = {{McLean}, I.~S.},
        title = "{Interpretation of the intrinsic polarization of early-type emission-line stars.}",
      journal = {\mnras},
         year = 1979,
        month = jan,
       volume = {186},
        pages = {265-285},
          doi = {10.1093/mnras/186.2.265},
       adsurl = {https://ui.adsabs.harvard.edu/abs/1979MNRAS.186..265M}
}

@ARTICLE{Nikolov+2023,
       author = {{Nikolov}, Y. and {Luna}, G.~J.~M. and {Stoyanov}, K.~A. and {Borisov}, G. and {Mukai}, K. and {Sokoloski}, J.~L. and {Avramova-Boncheva}, A.},
        title = "{Transient and asymmetric dust structures in the TeV-bright nova RS Oph revealed by spectropolarimetry}",
      journal = {\aap},
         year = 2023,
        month = nov,
       volume = {679},
          eid = {A150},
        pages = {A150},
          doi = {10.1051/0004-6361/202346997},
archivePrefix = {arXiv},
       eprint = {2309.11288},
 primaryClass = {astro-ph.SR},
       adsurl = {https://ui.adsabs.harvard.edu/abs/2023A&A...679A.150N}
}

@ARTICLE{Patterson+2022,
       author = {{Patterson}, Joseph and {Kemp}, Jonathan and {Monard}, Berto and {Myers}, Gordon and {de Miguel}, Enrique and {Hambsch}, Franz-Josef and {Warhurst}, Paul and {Rea}, Robert and {Dvorak}, Shawn and {Menzies}, Kenneth and {Vanmunster}, Tonny and {Roberts}, George and {Campbell}, Tut and {Starkey}, Donn and {Ulowetz}, Joseph and {Rock}, John and {Seargeant}, Jim and {Boardman}, James and {Lemay}, Damien and {Cejudo}, David and {Knigge}, Christian},
        title = "{IM Normae: The Death Spiral of a Cataclysmic Variable?}",
      journal = {\apj},
         year = 2022,
        month = jan,
       volume = {924},
       number = {1},
          eid = {27},
        pages = {27},
          doi = {10.3847/1538-4357/abec87},
       adsurl = {https://ui.adsabs.harvard.edu/abs/2022ApJ...924...27P}
}

@ARTICLE{Santamaria+2025,
       author = {{Santamar{\'\i}a}, E. and {Guerrero}, M.~A. and {Ramos-Larios}, G. and {Toal{\'a}}, J.~A. and {Sabin}, L.},
        title = "{A morphological catalogue of nova remnants}",
      journal = {\mnras},
         year = 2025,
        month = may,
       volume = {539},
       number = {1},
        pages = {246-264},
          doi = {10.1093/mnras/staf500},
archivePrefix = {arXiv},
       eprint = {2503.20053},
 primaryClass = {astro-ph.SR},
       adsurl = {https://ui.adsabs.harvard.edu/abs/2025MNRAS.539..246S}
}

@ARTICLE{Schaefer2010,
       author = {{Schaefer}, Bradley E.},
        title = "{Comprehensive Photometric Histories of All Known Galactic Recurrent Novae}",
      journal = {\apjs},
         year = 2010,
        month = apr,
       volume = {187},
       number = {2},
        pages = {275-373},
          doi = {10.1088/0067-0049/187/2/275},
archivePrefix = {arXiv},
       eprint = {0912.4426},
 primaryClass = {astro-ph.SR},
       adsurl = {https://ui.adsabs.harvard.edu/abs/2010ApJS..187..275S}
}

@ARTICLE{Schaefer+2010,
       author = {{Schaefer}, Bradley E. and {Pagnotta}, Ashley and {Shara}, Michael M.},
        title = "{The Nova Shell and Evolution of the Recurrent Nova T Pyxidis}",
      journal = {\apj},
         year = 2010,
        month = jan,
       volume = {708},
       number = {1},
        pages = {381-402},
          doi = {10.1088/0004-637X/708/1/381},
archivePrefix = {arXiv},
       eprint = {0906.0933},
 primaryClass = {astro-ph.SR},
       adsurl = {https://ui.adsabs.harvard.edu/abs/2010ApJ...708..381S}
}

@ARTICLE{Schulte-Ladbeck+1992,
       author = {{Schulte-Ladbeck}, R.~F. and {Nordsieck}, K.~H. and {Taylor}, M. and {Bjorkman}, K.~S. and {Magalhaes}, A.~M. and {Wolff}, M.~J.},
        title = "{The Wind Geometry of the Wolf-Rayet Star HD 191765}",
      journal = {\apj},
         year = 1992,
        month = mar,
       volume = {387},
        pages = {347},
          doi = {10.1086/171087},
       adsurl = {https://ui.adsabs.harvard.edu/abs/1992ApJ...387..347S}
}

@ARTICLE{Selvelli+2008,
       author = {{Selvelli}, P. and {Cassatella}, A. and {Gilmozzi}, R. and {Gonz{\'a}lez-Riestra}, R.},
        title = "{The secrets of T Pyxidis. II. A recurrent nova that will not become a SN Ia}",
      journal = {\aap},
         year = 2008,
        month = dec,
       volume = {492},
       number = {3},
        pages = {787-803},
          doi = {10.1051/0004-6361:200810678},
archivePrefix = {arXiv},
       eprint = {0904.1146},
 primaryClass = {astro-ph.SR},
       adsurl = {https://ui.adsabs.harvard.edu/abs/2008A&A...492..787S}
}

@ARTICLE{Shara+2018,
       author = {{Shara}, Michael M. and {Prialnik}, Dina and {Hillman}, Yael and {Kovetz}, Attay},
        title = "{The Masses and Accretion Rates of White Dwarfs in Classical and Recurrent Novae}",
      journal = {\apj},
         year = 2018,
        month = jun,
       volume = {860},
       number = {2},
          eid = {110},
        pages = {110},
          doi = {10.3847/1538-4357/aabfbd},
archivePrefix = {arXiv},
       eprint = {1804.06880},
 primaryClass = {astro-ph.SR},
       adsurl = {https://ui.adsabs.harvard.edu/abs/2018ApJ...860..110S}
}

@ARTICLE{Shore+2011,
       author = {{Shore}, S.~N. and {Augusteijn}, T. and {Ederoclite}, A. and {Uthas}, H.},
        title = "{The spectroscopic evolution of the recurrent nova T Pyxidis during its 2011 outburst. I. The optically thick phase and the origin of moving lines in novae}",
      journal = {\aap},
         year = 2011,
        month = sep,
       volume = {533},
          eid = {L8},
        pages = {L8},
          doi = {10.1051/0004-6361/201117721},
archivePrefix = {arXiv},
       eprint = {1108.3505},
 primaryClass = {astro-ph.SR},
       adsurl = {https://ui.adsabs.harvard.edu/abs/2011A&A...533L...8S}
}

@ARTICLE{Surina+2014,
       author = {{Surina}, F. and {Hounsell}, R.~A. and {Bode}, M.~F. and {Darnley}, M.~J. and {Harman}, D.~J. and {Walter}, F.~M.},
        title = "{A Detailed Photometric and Spectroscopic Study of the 2011 Outburst of the Recurrent Nova T Pyxidis from 0.8 to 250 Days after Discovery}",
      journal = {\aj},
         year = 2014,
        month = may,
       volume = {147},
       number = {5},
          eid = {107},
        pages = {107},
          doi = {10.1088/0004-6256/147/5/107},
archivePrefix = {arXiv},
       eprint = {1402.1109},
 primaryClass = {astro-ph.SR},
       adsurl = {https://ui.adsabs.harvard.edu/abs/2014AJ....147..107S}
}

@ARTICLE{Tanaka+2011,
       author = {{Tanaka}, Jumpei and {Nogami}, Daisaku and {Fujii}, Mitsugu and {Ayani}, Kazuya and {Kato}, Taichi},
        title = "{On the Rebrightenings of Classical Novae during the Early Phase}",
      journal = {\pasj},
         year = 2011,
        month = feb,
       volume = {63},
        pages = {159},
          doi = {10.1093/pasj/63.1.159},
archivePrefix = {arXiv},
       eprint = {1010.5611},
 primaryClass = {astro-ph.SR},
       adsurl = {https://ui.adsabs.harvard.edu/abs/2011PASJ...63..159T}
}

@ARTICLE{discovery_IAUC,
       author = {{Waagan}, E. and {Linnolt}, M. and {Pearce}, A.},
        title = "{T pyxidus.}",
      journal = {\iaucirc},
         year = 2011,
        month = jan,
       volume = {9205},
        pages = {1},
       adsurl = {https://ui.adsabs.harvard.edu/abs/2011IAUC.9205....1W}
}

@ARTICLE{Walter+2012,
       author = {{Walter}, Frederick M. and {Battisti}, Andrew and {Towers}, Sarah E. and {Bond}, Howard E. and {Stringfellow}, Guy S.},
        title = "{The Stony Brook/SMARTS Atlas of (mostly) Southern Novae}",
      journal = {\pasp},
         year = 2012,
        month = oct,
       volume = {124},
       number = {920},
        pages = {1057},
          doi = {10.1086/668404},
archivePrefix = {arXiv},
       eprint = {1209.1583},
 primaryClass = {astro-ph.SR},
       adsurl = {https://ui.adsabs.harvard.edu/abs/2012PASP..124.1057W}
}

@ARTICLE{Williams1992,
       author = {{Williams}, Robert E.},
        title = "{The Formation of Novae Spectra}",
      journal = {\aj},
         year = 1992,
        month = aug,
       volume = {104},
        pages = {725},
          doi = {10.1086/116268},
       adsurl = {https://ui.adsabs.harvard.edu/abs/1992AJ....104..725W}
}

@ARTICLE{Williams2012,
       author = {{Williams}, Robert},
        title = "{Origin of the ``He/N'' and ``Fe II'' Spectral Classes of Novae}",
      journal = {\aj},
         year = 2012,
        month = oct,
       volume = {144},
       number = {4},
          eid = {98},
        pages = {98},
          doi = {10.1088/0004-6256/144/4/98},
archivePrefix = {arXiv},
       eprint = {1208.0380},
 primaryClass = {astro-ph.SR},
       adsurl = {https://ui.adsabs.harvard.edu/abs/2012AJ....144...98W}
}



\appendix

\section{Instrumental and Interstellar Polarization Corrections}
\label{app:corrections}

The observed polarization results from the superposition of several contributions: the intrinsic polarization of the nova, the interstellar polarization (ISP), and the instrumental polarization. Let $Q_{\mathrm{obs}}$ and $U_{\mathrm{obs}}$ denote the observed Stokes parameters after data reduction:

\begin{align}
Q_{\mathrm{obs}} &= Q_{\mathrm{int}} + Q_{\mathrm{ISP}} + Q_{\mathrm{ins}}
\label{eq:Qobs}\\
U_{\mathrm{obs}} &= U_{\mathrm{int}} + U_{\mathrm{ISP}} + U_{\mathrm{ins}}
\label{eq:Uobs}
\end{align}

where $Q_{\mathrm{int}}$, $U_{\mathrm{int}}$ are the intrinsic Stokes parameters of the nova, $Q_{\mathrm{ISP}}$, $U_{\mathrm{ISP}}$ arise from interstellar polarization, and $Q_{\mathrm{ins}}$, $U_{\mathrm{ins}}$ are related to instrumental contamination.

\subsection{Instrumental Contamination}

\cite{Cikota+2017} investigated instrumental polarization induced by VLT/FORS2, including its wavelength dependence. They computed the weighted mean of Q and U as a function of wavelength using observations of unpolarized standard stars. A wavelength-dependent shift was detected, well described by first-order fits:

\begin{align}
Q_{\mathrm{ins}}(\lambda) &= \bigl[(9.66 \pm 1.04) \times 10^{-8}\bigr]\lambda 
                           + (3.29 \pm 6.34) \times 10^{-5}, \\
U_{\mathrm{ins}}(\lambda) &= \bigl[(7.28 \pm 0.90) \times 10^{-8}\bigr]\lambda 
                           - (4.54 \pm 0.55) \times 10^{-4}.
\end{align}

During the observation period of T Pyxidis, standard stars from the FORS1 Standard Fields and Stars were observed to perform polarimetric calibrations. The measured polarization matches the data from \cite{Cikota+2017}, so we adopted the same instrumental deviation.

\subsection{Interstellar Polarization}

The interstellar polarization can be considered as almost constant over the spectral range of interest, so its effect on a Q/U diagram is mostly to shift all points by a fixed vector. Consequently, the form of the Q/U characteristic features, such as loops, rotations, or depolarizations across spectral lines, remains unchanged. For this reason, many spectropolarimetric studies aiming to diagnose intrinsic asymmetries in the target object choose not to remove the interstellar component, since the excursions from the continuum that reveal geometrical structures are unaffected by this uniform translation in the Q/U plane.

However, because the transformation from $(Q,U)$ to $(P,\theta)$ is non-linear, the wavelength-dependent polarization degree $P(\lambda)$ and polarization angle $\theta(\lambda)$ calculated from the observed Stokes parameters (which include the interstellar component) may not accurately represent the intrinsic polarization state of the object. To estimate the ISP, we exploited the line depolarization effect, assuming that the Balmer recombination lines are intrinsically unpolarized and therefore depolarize the underlying continuum polarization (McLean 1979). Polarization detected at zero radial velocity (at the rest wavelength $\lambda_0$ of the Balmer lines) must be attributed to non-intrinsic contributions, namely ISP and instrumental effects:

\begin{align}
Q_{\mathrm{obs}}(\lambda_0) &= Q_{\mathrm{ISP}}(\lambda_0) + Q_{\mathrm{ins}}(\lambda_0), 
\label{eq:QobsRest}\\
U_{\mathrm{obs}}(\lambda_0) &= U_{\mathrm{ISP}}(\lambda_0) + U_{\mathrm{ins}}(\lambda_0).
\label{eq:UobsRest}
\end{align}

\begin{figure*}
    \centering
    \includegraphics[width=\textwidth]{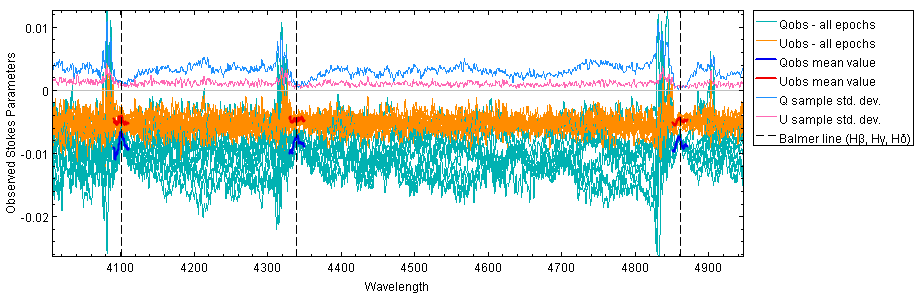}
    \caption{Multi-epoch distribution of the observed linear Stokes parameters over the observed spectral range (4000–5000 Å), including the Balmer recombination lines H$_\beta$, H$_\gamma$, and H$_\delta$. The individual $Q_{\mathrm{obs}}(\lambda)$ and $U_{\mathrm{obs}}(\lambda)$ spectra from all observing epochs are shown in cyan and orange, respectively. The dark blue and red curves indicate the mean values of $Q_{\mathrm{obs}}$ and $U_{\mathrm{obs}}$, computed within a narrow spectral window centred on the rest wavelength of each Balmer line. The light blue and pink curves show the corresponding wavelength-dependent standard deviations computed from all epochs. Vertical dashed lines mark the rest wavelengths of the Balmer transitions.}
\label{fig:StokesISP}
\end{figure*}

The minimal dispersion of the Stokes parameters within $\pm5\,\text{\AA}$ of H$_\beta$, H$_\gamma$ and H$_\delta$ (see Fig. \ref{fig:StokesISP})  confirms this assumption. Both the degree and angle of polarization are significantly non-zero and remain nearly constant with time, which is characteristic of the ISP.

The interstellar Stokes parameters $Q_{\mathrm{ISP}}(\lambda_0)$ and $U_{\mathrm{ISP}}(\lambda_0)$ can be calculated at the rest wavelength of the three Balmer lines using equations \eqref{eq:QobsRest} and \eqref{eq:UobsRest}, from which we can derive the degree and angle of the ISP at each Balmer line's rest wavelength using equations \eqref{eq:Degree} and \eqref{eq:Angle}. The resulting values are listed in Table~\ref{tab:ISP}.

\begin{table}
\centering
\caption{ISP values at the rest wavelengths of the Balmer lines.}
\small
\begin{tabular}{lcc}
\toprule
Line & $P_{\mathrm{ISP}}$ & $\theta_{\mathrm{ISP}}$ [deg] \\
\midrule
H$\beta$  & 0.00944 & 103.85 \\
H$\gamma$ & 0.00917 & 106.04 \\
H$\delta$ & 0.00866 & 106.66 \\
\bottomrule
\end{tabular}
\label{tab:ISP}
\end{table}

The best fit of the data to a Serkowski law is obtained with the parameters listed in Table~\ref{tab:SerkovParam}, so the degree of interstellar polarization can be expressed using the Serkowski law as: 

\begin{equation}
P_{\mathrm{ISP}}(\lambda) = 0.0097 \cdot 
\exp\!\left[-1.2 \, \ln^{2}\!\left(\frac{5500}{\lambda}\right)\right]
\label{eq:SerkDegree}
\end{equation}

\begin{table}
\centering
\caption{Best fit to a Serkowski law ($R^{2} = 0.93$)}
\small
\begin{tabular}{cc} 
\toprule
Serkowski law's parameter & Value \\
\midrule
$P_{\mathrm{max}}$   & 0.0097 \\
$\lambda_{\mathrm{max}}$   & 5500\text{\AA} \\
$K$   & 1.2 \\
\bottomrule
\end{tabular}
\label{tab:SerkovParam}
\end{table}

The small spectral dependence of the ISP angle within the observed spectral range visible in Table~\ref{tab:ISP} was approximated with a linear variation ($R^{2} = 0.989$):

\begin{equation}
\theta_{\mathrm{ISP}}(\lambda) = -7 \times 10^{-5}\,\lambda + 2.1336
\label{eq:SerkAngle}
\end{equation}

Equations \eqref{eq:SerkDegree} and \eqref{eq:SerkAngle} characterize the ISP through its polarization degree and position angle. The corresponding normalized Stokes parameters are obtained from

\begin{align}
Q = P\cos(2\theta)
\label{eq:Q=f(PT)}\\[1em]
U = P\sin(2\theta)
\label{eq:U=f(PT)}
\end{align}

The wavelength-dependent interstellar Stokes parameters are therefore

\begin{align}
Q_{\rm ISP}(\lambda)=P_{\rm ISP}(\lambda)
\cos\left[2\theta_{\rm ISP}(\lambda)\right] \\[1em]
U_{\rm ISP}(\lambda)=P_{\rm ISP}(\lambda)
\sin\left[2\theta_{\rm ISP}(\lambda)\right]
\end{align}

The intrinsic Stokes parameters are then obtained by subtracting both the interstellar and instrumental contributions from the observed Stokes parameters:

\begin{align}
Q_{\mathrm{int}}(\lambda) &= Q_{\mathrm{obs}}(\lambda) - Q_{\mathrm{ISP}}(\lambda) - Q_{\mathrm{ins}}(\lambda) \\
U_{\mathrm{int}}(\lambda) &= U_{\mathrm{obs}}(\lambda) - U_{\mathrm{ISP}}(\lambda) - U_{\mathrm{ins}}(\lambda)
\end{align}

Once this correction has been applied, the intrinsic polarization degree and position angle were subsequently recalculated from the corrected Stokes parameters, free from the systematic effects induced by interstellar polarization.

\begin{align}
P_{\mathrm{int}}(\lambda) &= \sqrt{\,Q_{\mathrm{int}}^{2}(\lambda) + U_{\mathrm{int}}^{2}(\lambda)} \\
\theta_{\rm int}(\lambda) &=\frac{1}{2}\operatorname{atan2}
\left[U_{\rm int}(\lambda),Q_{\rm int}(\lambda)\right]
\end{align}

\end{document}